\documentclass[sigconf, nonacm, natbib=false]{acmart}
\usepackage[font=small]{caption}
\usepackage{cleveref}
\usepackage{multirow}

\usepackage[T1]{fontenc}
\usepackage[utf8]{inputenc}
\usepackage{fancyvrb}
\usepackage{upquote}
\usepackage{xspace}
\usepackage{float}
\usepackage{tabularx}
\usepackage{booktabs}

\renewcommand{\sfdefault}{cmss}

\usepackage{tikz}
\usetikzlibrary{tikzmark}

\usepackage{xspace}
\usepackage{algorithm2e}
\SetKw{Continue}{continue}

\usepackage[normalem]{ulem}

\usepackage{comment}
\excludecomment{full}

\usepackage{subcaption}

\usepackage{pythonhighlight}

\usepackage [autostyle, english = american]{csquotes}
\MakeOuterQuote{"}
\newcommand{\techreport}[1]{#1}
\newcommand{\papertext}[1]{}

\newcommand\vldbdoi{XX.XX/XXX.XX}
\newcommand\vldbpages{XXX-XXX}
\newcommand\vldbvolume{14}
\newcommand\vldbissue{1}
\newcommand\vldbyear{2020}
\newcommand\vldbauthors{\authors}
\newcommand\vldbtitle{\shorttitle} 
\newcommand\vldbavailabilityurl{URL_TO_YOUR_ARTIFACTS}
\newcommand\vldbpagestyle{plain} 

\usepackage{hyperref}

\begin{document}

\newcommand{\topic}[1]{\vspace{-3.5pt}\smallskip \smallskip \noindent{\bf #1.}}

\newcommand{\mltrace}{\textsc{mltrace}\xspace}

\newcommand{\shreya}[1]{\textcolor{orange}{[Shreya: #1]}}
\newcommand{\aditya}[1]{\textcolor{purple}{[Aditya: #1]}}
\newcommand{\todo}[1]{\textcolor{blue}{TODO: #1}}

\newcommand{\mr}[1]{#1}
\newcommand{\rone}[1]{#1}
\newcommand{\rtwo}[1]{#1}
\newcommand{\rthree}[1]{#1}

\newcommand{\emtitle}[1]{\vspace{0.3em}\noindent{\em #1}}
\newcommand{\com}[1]{
\smallskip \noindent \textcolor{purple}{{\em #1}}}

\newenvironment{algo}{%
\renewenvironment{algocf}[1][h]{}{}% pass over the floating stuff
\algorithm
}{%
\endalgorithm
}

\makeatletter
\def\PY@reset{\let\PY@it=\relax \let\PY@bf=\relax%
    \let\PY@ul=\relax \let\PY@tc=\relax%
    \let\PY@bc=\relax \let\PY@ff=\relax}
\def\PY@tok#1{\csname PY@tok@#1\endcsname}
\def\PY@toks#1+{\ifx\relax#1\empty\else%
    \PY@tok{#1}\expandafter\PY@toks\fi}
\def\PY@do#1{\PY@bc{\PY@tc{\PY@ul{%
    \PY@it{\PY@bf{\PY@ff{#1}}}}}}}
\def\PY#1#2{\PY@reset\PY@toks#1+\relax+\PY@do{#2}}

\@namedef{PY@tok@w}{\def\PY@tc##1{\textcolor[rgb]{0.73,0.73,0.73}{##1}}}
\@namedef{PY@tok@c}{\let\PY@it=\textit\def\PY@tc##1{\textcolor[rgb]{0.25,0.50,0.50}{##1}}}
\@namedef{PY@tok@cp}{\def\PY@tc##1{\textcolor[rgb]{0.74,0.48,0.00}{##1}}}
\@namedef{PY@tok@k}{\let\PY@bf=\textbf\def\PY@tc##1{\textcolor[rgb]{0.00,0.50,0.00}{##1}}}
\@namedef{PY@tok@kp}{\def\PY@tc##1{\textcolor[rgb]{0.00,0.50,0.00}{##1}}}
\@namedef{PY@tok@kt}{\def\PY@tc##1{\textcolor[rgb]{0.69,0.00,0.25}{##1}}}
\@namedef{PY@tok@o}{\def\PY@tc##1{\textcolor[rgb]{0.40,0.40,0.40}{##1}}}
\@namedef{PY@tok@ow}{\let\PY@bf=\textbf\def\PY@tc##1{\textcolor[rgb]{0.67,0.13,1.00}{##1}}}
\@namedef{PY@tok@nb}{\def\PY@tc##1{\textcolor[rgb]{0.00,0.50,0.00}{##1}}}
\@namedef{PY@tok@nf}{\def\PY@tc##1{\textcolor[rgb]{0.00,0.00,1.00}{##1}}}
\@namedef{PY@tok@nc}{\let\PY@bf=\textbf\def\PY@tc##1{\textcolor[rgb]{0.00,0.00,1.00}{##1}}}
\@namedef{PY@tok@nn}{\let\PY@bf=\textbf\def\PY@tc##1{\textcolor[rgb]{0.00,0.00,1.00}{##1}}}
\@namedef{PY@tok@ne}{\let\PY@bf=\textbf\def\PY@tc##1{\textcolor[rgb]{0.82,0.25,0.23}{##1}}}
\@namedef{PY@tok@nv}{\def\PY@tc##1{\textcolor[rgb]{0.10,0.09,0.49}{##1}}}
\@namedef{PY@tok@no}{\def\PY@tc##1{\textcolor[rgb]{0.53,0.00,0.00}{##1}}}
\@namedef{PY@tok@nl}{\def\PY@tc##1{\textcolor[rgb]{0.63,0.63,0.00}{##1}}}
\@namedef{PY@tok@ni}{\let\PY@bf=\textbf\def\PY@tc##1{\textcolor[rgb]{0.60,0.60,0.60}{##1}}}
\@namedef{PY@tok@na}{\def\PY@tc##1{\textcolor[rgb]{0.49,0.56,0.16}{##1}}}
\@namedef{PY@tok@nt}{\let\PY@bf=\textbf\def\PY@tc##1{\textcolor[rgb]{0.00,0.50,0.00}{##1}}}
\@namedef{PY@tok@nd}{\def\PY@tc##1{\textcolor[rgb]{0.67,0.13,1.00}{##1}}}
\@namedef{PY@tok@s}{\def\PY@tc##1{\textcolor[rgb]{0.73,0.13,0.13}{##1}}}
\@namedef{PY@tok@sd}{\let\PY@it=\textit\def\PY@tc##1{\textcolor[rgb]{0.73,0.13,0.13}{##1}}}
\@namedef{PY@tok@si}{\let\PY@bf=\textbf\def\PY@tc##1{\textcolor[rgb]{0.73,0.40,0.53}{##1}}}
\@namedef{PY@tok@se}{\let\PY@bf=\textbf\def\PY@tc##1{\textcolor[rgb]{0.73,0.40,0.13}{##1}}}
\@namedef{PY@tok@sr}{\def\PY@tc##1{\textcolor[rgb]{0.73,0.40,0.53}{##1}}}
\@namedef{PY@tok@ss}{\def\PY@tc##1{\textcolor[rgb]{0.10,0.09,0.49}{##1}}}
\@namedef{PY@tok@sx}{\def\PY@tc##1{\textcolor[rgb]{0.00,0.50,0.00}{##1}}}
\@namedef{PY@tok@m}{\def\PY@tc##1{\textcolor[rgb]{0.40,0.40,0.40}{##1}}}
\@namedef{PY@tok@gh}{\let\PY@bf=\textbf\def\PY@tc##1{\textcolor[rgb]{0.00,0.00,0.50}{##1}}}
\@namedef{PY@tok@gu}{\let\PY@bf=\textbf\def\PY@tc##1{\textcolor[rgb]{0.50,0.00,0.50}{##1}}}
\@namedef{PY@tok@gd}{\def\PY@tc##1{\textcolor[rgb]{0.63,0.00,0.00}{##1}}}
\@namedef{PY@tok@gi}{\def\PY@tc##1{\textcolor[rgb]{0.00,0.63,0.00}{##1}}}
\@namedef{PY@tok@gr}{\def\PY@tc##1{\textcolor[rgb]{1.00,0.00,0.00}{##1}}}
\@namedef{PY@tok@ge}{\let\PY@it=\textit}
\@namedef{PY@tok@gs}{\let\PY@bf=\textbf}
\@namedef{PY@tok@gp}{\let\PY@bf=\textbf\def\PY@tc##1{\textcolor[rgb]{0.00,0.00,0.50}{##1}}}
\@namedef{PY@tok@go}{\def\PY@tc##1{\textcolor[rgb]{0.53,0.53,0.53}{##1}}}
\@namedef{PY@tok@gt}{\def\PY@tc##1{\textcolor[rgb]{0.00,0.27,0.87}{##1}}}
\@namedef{PY@tok@err}{\def\PY@bc##1{{\setlength{\fboxsep}{\string -\fboxrule}\fcolorbox[rgb]{1.00,0.00,0.00}{1,1,1}{\strut ##1}}}}
\@namedef{PY@tok@kc}{\let\PY@bf=\textbf\def\PY@tc##1{\textcolor[rgb]{0.00,0.50,0.00}{##1}}}
\@namedef{PY@tok@kd}{\let\PY@bf=\textbf\def\PY@tc##1{\textcolor[rgb]{0.00,0.50,0.00}{##1}}}
\@namedef{PY@tok@kn}{\let\PY@bf=\textbf\def\PY@tc##1{\textcolor[rgb]{0.00,0.50,0.00}{##1}}}
\@namedef{PY@tok@kr}{\let\PY@bf=\textbf\def\PY@tc##1{\textcolor[rgb]{0.00,0.50,0.00}{##1}}}
\@namedef{PY@tok@bp}{\def\PY@tc##1{\textcolor[rgb]{0.00,0.50,0.00}{##1}}}
\@namedef{PY@tok@fm}{\def\PY@tc##1{\textcolor[rgb]{0.00,0.00,1.00}{##1}}}
\@namedef{PY@tok@vc}{\def\PY@tc##1{\textcolor[rgb]{0.10,0.09,0.49}{##1}}}
\@namedef{PY@tok@vg}{\def\PY@tc##1{\textcolor[rgb]{0.10,0.09,0.49}{##1}}}
\@namedef{PY@tok@vi}{\def\PY@tc##1{\textcolor[rgb]{0.10,0.09,0.49}{##1}}}
\@namedef{PY@tok@vm}{\def\PY@tc##1{\textcolor[rgb]{0.10,0.09,0.49}{##1}}}
\@namedef{PY@tok@sa}{\def\PY@tc##1{\textcolor[rgb]{0.73,0.13,0.13}{##1}}}
\@namedef{PY@tok@sb}{\def\PY@tc##1{\textcolor[rgb]{0.73,0.13,0.13}{##1}}}
\@namedef{PY@tok@sc}{\def\PY@tc##1{\textcolor[rgb]{0.73,0.13,0.13}{##1}}}
\@namedef{PY@tok@dl}{\def\PY@tc##1{\textcolor[rgb]{0.73,0.13,0.13}{##1}}}
\@namedef{PY@tok@s2}{\def\PY@tc##1{\textcolor[rgb]{0.73,0.13,0.13}{##1}}}
\@namedef{PY@tok@sh}{\def\PY@tc##1{\textcolor[rgb]{0.73,0.13,0.13}{##1}}}
\@namedef{PY@tok@s1}{\def\PY@tc##1{\textcolor[rgb]{0.73,0.13,0.13}{##1}}}
\@namedef{PY@tok@mb}{\def\PY@tc##1{\textcolor[rgb]{0.40,0.40,0.40}{##1}}}
\@namedef{PY@tok@mf}{\def\PY@tc##1{\textcolor[rgb]{0.40,0.40,0.40}{##1}}}
\@namedef{PY@tok@mh}{\def\PY@tc##1{\textcolor[rgb]{0.40,0.40,0.40}{##1}}}
\@namedef{PY@tok@mi}{\def\PY@tc##1{\textcolor[rgb]{0.40,0.40,0.40}{##1}}}
\@namedef{PY@tok@il}{\def\PY@tc##1{\textcolor[rgb]{0.40,0.40,0.40}{##1}}}
\@namedef{PY@tok@mo}{\def\PY@tc##1{\textcolor[rgb]{0.40,0.40,0.40}{##1}}}
\@namedef{PY@tok@ch}{\let\PY@it=\textit\def\PY@tc##1{\textcolor[rgb]{0.25,0.50,0.50}{##1}}}
\@namedef{PY@tok@cm}{\let\PY@it=\textit\def\PY@tc##1{\textcolor[rgb]{0.25,0.50,0.50}{##1}}}
\@namedef{PY@tok@cpf}{\let\PY@it=\textit\def\PY@tc##1{\textcolor[rgb]{0.25,0.50,0.50}{##1}}}
\@namedef{PY@tok@c1}{\let\PY@it=\textit\def\PY@tc##1{\textcolor[rgb]{0.25,0.50,0.50}{##1}}}
\@namedef{PY@tok@cs}{\let\PY@it=\textit\def\PY@tc##1{\textcolor[rgb]{0.25,0.50,0.50}{##1}}}

\def\PYZbs{\char`\\}
\def\PYZus{\char`\_}
\def\PYZob{\char`\{}
\def\PYZcb{\char`\}}
\def\PYZca{\char`\^}
\def\PYZam{\char`\&}
\def\PYZlt{\char`\<}
\def\PYZgt{\char`\>}
\def\PYZsh{\char`\#}
\def\PYZpc{\char`\%}
\def\PYZdl{\char`\$}
\def\PYZhy{\char`\-}
\def\PYZsq{\char`\'}
\def\PYZdq{\char`\"}
\def\PYZti{\char`\~}
% for compatibility with earlier versions
\def\PYZat{@}
\def\PYZlb{[}
\def\PYZrb{]}
\makeatother

% \ifrebuttal
% \twocolumn
% \input{rebuttal.tex}
% \newpage
% \fi

\title{Towards Observability for \\ \mr{Production} Machine Learning Pipelines}
\subtitle{[Vision Paper]}


\author{Shreya Shankar}
\affiliation{%
  \institution{UC Berkeley}
}
\email{shreyashankar@berkeley.edu}

\author{Aditya G. Parameswaran}
\affiliation{%
  \institution{UC Berkeley}
}
\email{adityagp@berkeley.edu}

\begin{abstract}
\small
Software organizations are increasingly incorporating machine learning (ML) into their product offerings, driving a need for new data management tools. Many of these tools facilitate the initial development of ML applications, but sustaining these applications post-deployment is difficult due to lack of real-time feedback (i.e., labels) for predictions and silent failures that could occur at any \mr{component} of the ML pipeline (e.g., data distribution shift \mr{or anomalous features}). We propose a new type of data management system that offers end-to-end \emph{observability}, or visibility into complex system behavior, for \mr{deployed} ML pipelines through assisted (1) detection, (2) diagnosis, and (3) reaction to ML-related bugs. We describe new research challenges and suggest preliminary solution ideas in all three aspects. Finally, we introduce an example architecture for a "bolt-on" ML observability system, or one that wraps around existing tools in the stack.
\end{abstract}

\maketitle

\section{Introduction}

Organizations are devoting increasingly more resources towards developing and deploying applications powered by machine learning (ML). ML applications rely on pipelines that span multiple heterogeneous stages or \emph{components}, such as feature generation and model training, requiring specialized data management tools. Most work in data management for ML concentrates on specific components, e.g.,
preprocessing~\cite{mlinspect, dagger}, 
or model
training~\cite{modeldb, mistique,modelhub,garcia2020hindsight}. 
Additionally, some industry solutions 
have garnered widespread adoption by handling 
data management issues that stem from experimenting with models~\cite{mlflow,wandb}. 

% These components each come with their own challenges, 
% such as data preprocessing issues~\cite{schelter,dde,dagger}, 
% dealing with covariate shift~\cite{sugiyama, dmc},
% prediction serving problems~\cite{velox}, 
% and periodic retraining complications~\cite{prodml}.  

% To support visibility into specific components of the ML pipeline,
% the database 
% community has proposed a variety of solutions,
% e.g., for identifying data bugs during 
% preprocessing~\cite{mlinspect, dagger},
% and for logging models and model metadata during training 
% for post-hoc debugging~\cite{modeldb, mistique,modelhub,garcia2020hindsight}.
% Additionally, industry solutions such as 
% MLFlow~\cite{mlflow} and Weights \& Biases~\cite{wandb} 
% have garnered widespread adoption by handling 
% data management issues that stem from experimenting with large numbers of models. As a result of all of these component-centric data management tools, building an ML pipeline has never been easier.

However, there are many unaddressed challenges in \emph{sustaining} ML pipelines \mr{once built}: maintaining, debugging, and improving them after the initial deployment. Various best practices for "production ML" and failure case studies highlight the dire need for ML sustainability~\cite{rubric, mldebt}. We posit that for sustainability, ML practitioners should be able to (1) detect, (2) diagnose, and (3) react to bugs post-deployment. Compared to traditional software systems, which typically only break when there are infrastructure issues, ML pipelines can also fail unpredictably due to data issues---and therefore are uniquely challenging to sustain in all three aspects.

\vspace{-1.5pt}
\topic{Bug Detection: Hard Due to Feedback Delays} 
It is well-known that data distributions change 
or shift over time, causing model performance to drop~\cite{sugiyama, dmc}. 
Detecting performance drops post-deployment 
is challenging due to lack of "ground-truth" data: 
in many production ML systems, 
feedback on predictions, or labels, 
can arrive at a later time. 
Furthermore, in many pipelines, 
only a few labels arrive (e.g., labelers manually annotate some predictions, 
or only a handful of predicted outputs are displayed to the user). 
As a result, practitioners are unable to monitor simple ML 
metrics such as accuracy in real-time. 
As an alternative, end-to-end ML frameworks such as TFX~\cite{tfx} and Sagemaker~\cite{sagemaker} 
monitor internal pipeline 
state or health via distance metrics~\cite{massey1951kolmogorov}
over distributions of ML features 
and outputs over time. 
These proxies often produce too many false positives 
and thus do not accurately determine 
when models are underperforming, 
as we will discuss further in Section~\ref{sec:related}.

\topic{Bug Diagnosis: Hard Due to Pipeline Complexity} 
Even if a failure is confidently detected, the complex, 
highly intertwined nature of components 
in the ML pipeline 
makes it hard to \mr{understand where bugs could lie}. 
For \mr{production} ML pipelines, 
"changing anything changes everything (CACE)," 
causing predictions 
to vary unpredictably~\cite{mldebt}. 
\mr{Consequently, production} ML uniquely 
suffers from silent \mr{pipeline bugs, such as corrupted or stale subsets of features. Practitioners painstakingly enumerate and maintain (i.e., tune) data quality constraints for component inputs and outputs~\cite{dataval, deequ}, motivating automatic specification and maintenance of precise constraints at the component level.} 
% For example, changing data cleaning criteria (e.g., upper and lower bounds for a 
% column) might change the feature and prediction distributions. 
% Moreover, models are periodically retrained 
% and redeployed over time~\cite{prodml}, 
% making debugging a nightmare 
% if practitioners do not log, version, 
% and track the lineage of every artifact 
% generated by every component in the 
% pipeline. 
% Finally, 
% (i.e., low-quality predictions are generated 
% even when there are bugs). 
% Consequently, failures in different components 
% can result in the same output: 
% for example, both a broken sensor 
% that produces raw data and 
% an incorrect join in the feature 
% generation component can yield 
% too many null values for a column. 
% This motivates fine-grained logging of inputs and outputs at the component level.

\topic{Bug Fixes: Hard Due to No Obviously Correct Answers} 
Even if users 
can successfully 
\mr{pinpoint all pipeline bugs}, 
there can be many ways to bring 
model performance back up to a desirable level, 
and effectiveness depends on the 
nature of the data or task. 
For example, \mr{different components, when fixed, can cause different magnitudes of improvement in ML performance}. 
Users often have no sense of \mr{what to fix first}, relative to the costs
in resources and time.

\topic{ML Observability}
The challenges outlined above motivate the need for \emph{observability}~\cite{sridharan2018distributed}, or ``{\em better visibility into understanding the complex
behavior of software using telemetry collected ... at run time''}~\cite{karumuri2021towards}, tailored for ML pipelines. 
Observability encompasses more than just 
monitoring predefined metrics 
that capture holistic system health\techreport{ (i.e., known-unknowns)}---it 
also allows practitioners to ask questions about 
how systems behaved on historical outputs\techreport{ (i.e., unknown-unknowns)}, 
or perform "needle-in-a-haystack" queries\techreport{. 
The north star for software observability systems 
is to give users the power to ask new questions 
of historical system behavior without gathering new data}~\cite{majors_2021}.

\topic{Contributions}  In this paper, we discuss unaddressed research challenges in ML observability as a call-to-arms for the database community to contribute to this nascent research
direction. We propose the concept of a "bolt-on" observability system for ML pipelines---one that does not require users to rewrite all their code to use a specific framework.
ML application developers assemble their pipelines in an ad-hoc manner
employing a myriad of tools along the way, and our bolt-on observability
system must interoperate with such heterogeneous pipelines.
For example, practitioners may use a Hive metastore to catalog 
raw data~\cite{hive}, Deequ for data validation~\cite{deequ}, and Weights \& Biases for experiment tracking~\cite{wandb}.

For our bolt-on observability system to 
address bug detection, diagnosis, and fixing needs, 
we propose a three-pronged approach:
% \smallskip

\noindent (1) Monitoring approximations of top-line, i.e., business-critical, ML metrics to alert users of ML performance drops even when there may not be real-time labels. In Section~\ref{sec:coarsegrained}, we propose automated techniques that rely on lightweight proxies to bin predictions and estimate metrics based on importance weighting, drawing on the approximate query processing and streaming literature. \\
(2) \mr{Given ML performance drops, identifying issues in inputs and outputs for each component in the pipeline to aid diagnosis. In Section~\ref{sec:finegrained}, we propose logging fine-grained information across provenance snapshots, automatically specifying and tuning data quality constraints,} and adversarially learning differences between training and live data to track distribution shift. \\ 
(3) \mr{Tracing ML bugs back to silent data and engineering-related issues (i.e., pipeline bugs). In Section~\ref{sec:reacting}, we describe tracking feedback delays and column-wide error scores across dataflow graphs to assist practitioners in repairing broken components.} 

% \agp{Flip this to talk about diagnosis first -- in terms of identifying components with issues at a given point in time, given that there is a performance drop via (1) -- then provenance can come second} \shreya{done}

% \agp{Would be good if we can emphasize our roadmap a bit: the metrics - single component/time - cross component/time, as well as a spectrum of errors in each case}

% \smallskip
%\end{enumerate}

% \shreya{flush this out}
% \begin{enumerate}
%     \item Monitoring coarse-grained metrics to help users detect when ML pipelines fail
%     \item Logging fine-grained metrics for users to query while diagnosing ML failures
%     \item Suggesting strategies for users to react to ML failures
% \end{enumerate}
% \agp{Annotating each of the bullets above with the Section numebr would be good.} \shreya{done}

\mr{In Section~\ref{sec:researchprobs}, we present a roadmap (\Cref{fig:roadmap}) of challenges and preliminary solutions---\emph{detecting} drops in holistic ML metrics (e.g., accuracy), \emph{diagnosing} them by tracking point-in-time, component-level issues, and \emph{reacting} to the drops by analyzing cross-time, cross-component issues.} In Section~\ref{sec:system}, we discuss an example of a bolt-on ML observability system architecture and introduce our vision for \mltrace, a lightweight bolt-on ML observability tool, which has
already received preliminary interest
from practitioners with over \rtwo{400} GitHub stars (\texttt{github.com/loglabs/mltrace}).

\begin{figure}
 \papertext{\vspace{-12pt}}
    \centering
    \includegraphics[width=0.7\linewidth]{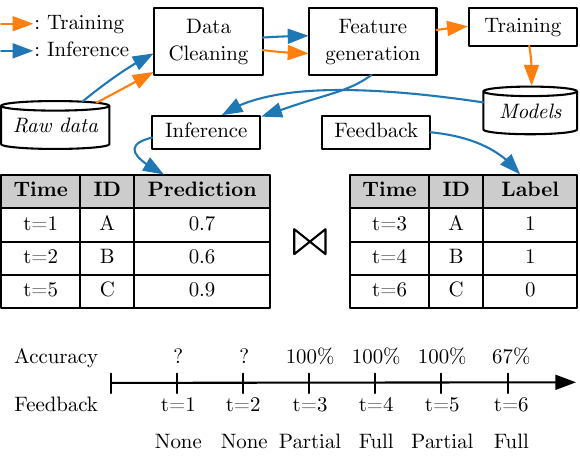}
    \papertext{\vspace{-15pt}}
    \caption{A generic end-to-end ML pipeline. \techreport{The inference component generates predictions and the feedback component produces labels.} Feedback comes with delay, impacting real-time accuracy.}
    \label{fig:tradpipeline}
    \papertext{\vspace{-21pt}}
\end{figure}

% \subsection{ML Observability}
\begin{full}
Software observability~\cite{sridharan2018distributed} is a concept drawn from control theory,
and refers to bringing  ``{\em better visibility into understanding the complex
behavior of software using telemetry collected ... at run time''}~\cite{karumuri2021towards}. 
Existing software observability tools 
lack support for ML pipelines, 
which require many additional performance metrics 
and frequent "snapshotting" due to ever-changing 
data flow through the pipeline, 
posing new research challenges in data management. 

Typical software observability solutions 
present metadata in the form of \emph{logs}, \emph{metrics}, 
and \emph{traces}\techreport{\footnote{\scriptsize{\em Events} is another form of software observability artifact; however
they are less relevant in an ML setting~\cite{karumuri2021towards}. These artifacts are together referred to as MELT.}} 
and synthesize this information to 
allow engineers to ask questions about system health.
ML applications add new criteria to system health, 
as performance extends beyond system uptime; 
it also relies on the quality of predictions 
or outputs. 
We use the three aforementioned pillars of software observability 
as inspiration for the key facets of our ML observability approach, and outline the corresponding research challenges, next\footnote{\scriptsize Logs, metrics, and traces,
as used in software observability, represent different debugging patterns (i.e., finding a "needle in a haystack," inspecting the cause of an alert, and understanding the lifecycle of an action respectively). 
These debugging patterns and corresponding metadata 
do not map directly
to ML observability, motivating our slight modification.}.

\topic{Logging} 
Traditional software applications record events
through logs, or event-related metadata produced
at runtime, such as stack traces and error codes for practitioners to search while debugging to answer their "needle-in-a-haystack" queries.
ML applications add complexity to logging
since they can experience silent failures 
(e.g., covariate shift or concept drift),
which may not be reflected in typical output or error logging statements.
Thus, our approach must capture inputs and outputs
of intermediate components, both of which can be computationally
expensive and bulky to store (e.g., large DNN models).
Additionally, it must be able to identify how inputs and outputs
feed into each other across the pipeline, despite pipeline
components being run at various frequencies, e.g.,
the training step in Figure~\ref{fig:tradpipeline} being run weeks prior to inference.
It should also support this logging in a lightweight manner
with minimal user input.

\topic{Monitoring} 
In traditional software applications, 
engineers monitor numerical metrics 
that directly represent how well 
an application is performing (i.e., mean response time) 
and receive alerts when these metric values exceed 
a predefined threshold. 
Such metrics are typically simple 
to measure, as the values can easily be 
captured in the application and 
aggregations are straightforward. 
Monitoring is not straightforward in ML
because the success of an ML application 
depends on model performance, 
which cannot easily be measured;
labels, or true values, are often 
not provided in real-time in the production endpoint. 
Even if direct feedback 
is available (e.g., when a user clicks on an ad), prediction and feedback streams are can be separate, making it expensive to compute
ML metrics, such as F1 and t-test scores, at scale.
Our approach must support a variety of ML-centric
metrics efficiently in a plug-and-play fashion 
without the user having to write a lot of code,
with these metrics being tracked in-situ as components
of the pipelines are run repeatedly. 
Users should also be able to set alerts, triggers,
or constraints to ensure overall ML pipeline health.

\topic{Querying} 
To debug systems from a holistic, end-to-end perspective, 
users need to be able to query the logs and metrics produced by an observability tool. To debug traditional software applications, engineers typically first inspect a trace, 
or the end-to-end journey of a data point, 
to understand the lineage of that request 
and determine where the bug may lie. 
This lineage may be straightforward to track 
when transformations are done in a single framework (e.g., REST), 
but ML applications are built 
using heterogeneous stacks of many tools. Having a complicated technical stack fragments access to information about data flow and provenance. 
Our approach must therefore integrate 
with a heterogeneous set of 
tools and still support lineage tracking.
Second, ML bugs are rarely constrained to a single output, as ML pipelines will not have 100\% accuracy. Practitioners typically investigate errors belonging to a group of outputs, or a \emph{slice}.

Our approach must support slice-based lineage querying where slices could be any subgroup defined on-demand, efficiently. Third, in many regulated industries such as finance or law, users may need to query over previous months or even years, requiring data to be retained over unusually long periods of time. This differs from traditional software applications, which commonly retain metadata on the order of days or weeks.
\end{full}

\papertext{\vspace{-6pt}}
\section{Background}
\label{sec:related}
% General themes in existing work:

% \begin{itemize}
%     \item People invent new abstractions, which is complicated and a barrier to adoption
%     \item People that invent logging solutions tend to invent their own query language, which is also a barrier to adoption
%     \item Generally focused on a part of the pipeline (like preprocessing or training)
%     \item End-to-end solutions lock you into a platform / ecosystem (ex: TFX) or are heavy because they require platform-specific integrations (ex: mlflow and modeldb have separate integrations with sklearn, tf, pytorch, etc)
% \end{itemize}

% Separate related work among different axes:

% \begin{itemize}
%     \item dev vs prod
%     \item fine grained vs coarse grained (component-specific vs end-to-end)
%     \item single-user vs multi-user (how collaborative?)
%     \item user effort (integration effort, interaction effort)
%     \item computational overhead
%     \item Function
%     \begin{itemize}
%         \item Metric monitoring over time
%         \item Logging (artifact management)
%         \item Tracing (provenance, history of an output)
%     \end{itemize}
% \end{itemize}
% \vspace{-1em}
We discuss prior work in data management for ML pipelines and current end-to-end ML pipeline frameworks.

\noindent \textbf{ETL, Assertions, and Experiment Tracking.} \mr{Input data for ML models is typically constructed through a series of ETL workloads. Faulty predictions can stem from such workloads, such as incorrectly performing missing value imputation~\cite{schelter}. Tools like Dagger~\cite{dagger} and mlinspect~\cite{mlinspect} help practitioners detect data-related bugs in ML preprocessing components. Our focus is instead on bugs that arise post-deployment.} Other tools~\cite{mlflow,modeldb,wandb} focus on experiment tracking, one of the biggest pain points in generating models for production ML pipelines. \mr{However, none of these tools determine {\em if} and \emph{why} production pipelines are failing}. \\
\textbf{Data Quality Assertions.} \rone{Other work~\cite{deequ, greatexp, tfx, Kang2020ModelAF} proposes libraries of 
assertions to 
be embedded in ML code, however, they provide no guidance for 
which assertions to embed.}
\techreport{Since these assertions are 
often written as part of a main application, 
they may not be easily reusable across pipelines.}
\rthree{Additionally, results of these 
tests must be externally 
logged with a separate service for users to query post-hoc. 
While data quality assertions are certainly 
valuable for catching egregious issues (e.g., negative values for columns that should be positive), ML pipeline performance can drop over time without failing user-embedded assertions}. \\
\textbf{Detecting data shift.}
\label{sec:detectingdatashift}
Many papers in the ML literature 
discuss how various forms 
of data shift\techreport{ (e.g., concept shift, 
covariate shift, prior probability shift)} 
cause model performance to degrade~\cite{sugiyama, unifyingview, bbse}. 
To address such shift problems, 
the ML community has proposed monitoring 
distance metrics across distributions 
of features and predictions\techreport{, 
such as the Kolmogorov-Smirnov (K-S) test statistic for numerical features or dimensionality-reduced features and 
the Chi-Squared test statistic for categorical features}~\cite{failingloudly}. 
However, 
with thousands of features and seasonal changes in data, such methods may not correctly flag shift, might trigger too many alarms and cause alert "fatigue"\techreport{ or result in confusion (e.g., the K-S test statistic is significant for one feature but not another)}~\cite{dataval}. 
Thus, there is a need for higher-precision methods that detect, diagnose, and react to data shift.
\rone{Some research in post-deployment ML debugging focuses on finding slices (i.e., predicates) where models perform poorly~\cite{slicefinder, sliceline}, but these methods require labels, which are not always available.} 

\noindent \textbf{Unresolved Observability Challenges in Existing Tools.}
End-to-end frameworks such as Sagemaker~\cite{sagemaker} and TFX~\cite{tfx} provide logging at the component level but only support primitive monitoring \rone{based on user-specified metrics and similarly do not help address data shift}. 
These frameworks also force their users 
to rewrite their pipeline using their DSLs. 
\techreport{For example, to use TFX, 
users must write their data 
processing pipelines using Apache Beam, 
manipulate data with TFData, build 
models in Tensorflow, and serve models via Tensorflow Serving.}
To avoid having users perform a cumbersome rewrite, some proprietary tools only monitor 
\mr{features and} predictions\techreport{ through an API, 
which cannot flag all problems 
or suggest where problems
lie in the pipeline because they lack end-to-end visibility}. 
Other declarative frameworks~\cite{overton, molino2019ludwig} allow users
to declaratively specify end-to-end ML pipelines \mr{ without supporting the identification of deployment bugs}. 
\techreport{ 
Moreover, ML practitioners often prefer to use their homegrown hodgepodge of tools rather than rewrite their code in a separate framework. Thus,
we advocate for an observability solution that can interoperate
with such tools.}

\begin{full}
\section{Motivating example}
\label{sec:motivate}
Consider a fictitious, small ride-sharing startup called Rebu that wants to predict which riders might give their drivers a tip and release this feature to their most loyal drivers. A product manager works with a team of data scientists and engineers to implement an endpoint that returns the probability that the rider will give a tip. This endpoint will be handed off to a front-end engineer at Rebu. For context, we illustrate the high-level architecture of an end-to-end production machine learning pipeline in figure \ref{fig:tradpipeline}.

\begin{figure*}
    \centering
    \includegraphics[width=\textwidth]{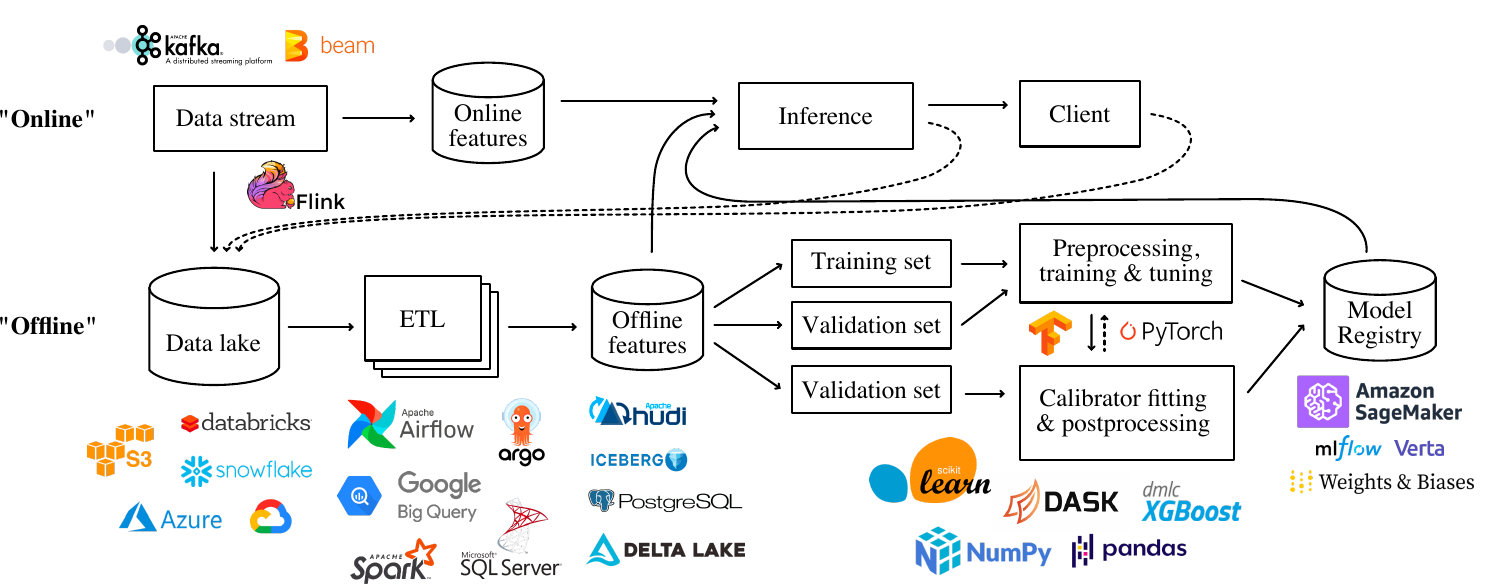}
    \caption{High-level architecture of a generic end-to-end machine learning pipeline.}
    \label{fig:tradpipeline}
\end{figure*}

\subsection{Phase 1: ML experimentation}

Suppose the data science team at Rebu has several tools at their disposal, including but not limited to:

\begin{itemize}
    \item Tables of users' data regularly and thanklessly refreshed by some underpaid data engineers
    \item Hosted Jupyter \cite{jupyter} programming environment to experiment with modeling ideas in any libraries of their choice
    \item Tools such as MLFlow \cite{mlflow} or Weights \& Biases \cite{wandb} to manage artifacts generated in their experiments
\end{itemize}

They perform some ETL workloads on the tables of data to derive some features, split the features into train and validation sets, and train an \texttt{xgboost} \cite{xgboost} model that achieves high F1 scores on the validation sets. They clean up the code in their Jupyter notebooks and make a pull request in the data science code repository with featurization code, sample inference code, and the MLFlow model URI.

\subsection{Phase 2: Productionizing the ML pipeline}

An ML engineer reviews the pull request with the goal of putting this project in production. She skims the featurization code, written in Python, in the pull request to find that some features require "dynamic" data, such as current location, and other features require more "static" data, such as when the rider signed up for Rebu. She then writes a Dask pipeline to compute the static features and write them to a table named \texttt{dim\_\_tip\_prediction\_features} and another cron-scheduled Dask pipeline to infrequently compute some dynamic features that don't depend on the live data, such as the average tip the rider gave in the past, and write them to a table named \texttt{fct\_\_tip\_prediction\_features}. Both tables are versioned with a partition that stores the timestamp when the records were added to the table. To finish the featurization step, the ML engineer writes a pipeline in SparkSQL \cite{spark} to compute dynamic features on Rebu's live Kafka streams \cite{kafka}.

Now, the ML engineer writes the inference endpoint to tie all of the steps together. In this endpoint, she calls the SparkSQL pipeline and performs joins on the \texttt{dim} and \texttt{fct} feature tables to get a vector of features to pass to the model. She then loads the model, runs inference on the features, and returns the probability the user will give a tip. Finally, she writes some crude tests to ensure the end-to-end pipeline runs successfully on valid data and fails gracefully when data might be missing or the Dask pipelines miss their SLAs. Another software engineer reviews the code and pushes it to production, creating some basic metrics to add to their Prometheus monitoring system such as response time, number of requests, and average prediction value.

\subsection{Phase 3: Post-deployment}

Once the end to end pipeline is in production, people have new responsibilities. Some ML engineers rotate being "on-call" in case the pipeline is down. Some data scientists try new ideas to improve the model. An ambitious ML engineer intern joins the team, eager to get a taste for ML in the "real world." In the first half of the internship, they write Python code to retrain the model every few days and promote the newest models to production with MLFlow. In the second half of the internship, the intern adds some new "windowed" features that compute aggregates over the past month, which boosts the model's performance. The intern leaves and the holiday season begins. The ML engineer who wrote the initial production pipeline takes parental leave for the upcoming Q1.

The team's product manager returns from their holiday break to find some forwarded angry emails from Very Important Drivers saying their riders are tipping less than they used to. In the team meeting, the on-call engineer says the mean prediction value is higher than usual, but so is the actual average tip value over the holiday season. The data scientists point out that the models are retrained frequently, so data drift issues might not be as pronounced. A data engineer reports that all pipelines met their SLAs. A software engineer who took a Coursera course on machine learning offers to debug the incorrect predictions, but he quickly finds himself in a black hole.

\subsection{Phase 4: Debugging faulty outputs from the ML pipeline}

The first question that the software engineer asks is: what was involved in producing these outputs? He is shocked to find that there is no tool to trace these outputs. With some help from a data analyst who knows all the tables, they construct a new table of properly dated outputs and ground truth tip values for the Very Important Drivers who complained. The predictions are high, but the tip values are low. The software engineer learns that, given the dates of these predictions, different model binaries are used in making these predictions. He painstakingly finds all the model binaries used and checks their MLFlow logs to verify that their train and validation scores were high. They check the Git repository history in an attempt to find a code change that may have caused the performance drop, but there are too many commits.

A week later, after reconstructing end-to-end traces for each prediction and each piece of data involved in making the predictions, the engineer shows up to the next all-hands meeting with a long list of problems with the pipeline. They showcase a bug introduced when the intern added new windowed features, as some of the new features accidentally shared the same name as some of the old features, unintentionally amplifying their weight in model training and inference. They also mention that the \texttt{dim} feature table was not derived from the freshest source of data at the time of creation. As they discuss more issues, it becomes painfully clear that the team is poorly equipped to handle their production ML pipeline. They cannot easily triage issues that happened in the past, let alone anticipate when things will go wrong in the future.

\end{full}

\begin{full}
\section{Observability requirements}
Inspired by the software industry's focus on supporting both \emph{retrospective} and \emph{proactive} queries in observability tools, we define notions of retrospectivity and proactivity in ML pipeline observability.

\subsection{Definitions}

\begin{definition}[Retrospectivity]
A query is retrospective if it is inquiring about the pipeline's behavior with respect to outputs that are known to be faulty.
\end{definition}

The example in section \ref{sec:motivate} motivates retrospective queries, since some faulty predictions were known before the engineer began debugging. Some retrospective queries that relate specifically to ML pipelines include:

\begin{itemize}
    \item What were the training sets used in any models involved in making a specific prediction?
    \item What is the historical performance over time for a specified subgroup, such as the predictions made for a single driver's rides over the last month?
\end{itemize}

\begin{definition}
A query is proactive if it is inquiring about the pipeline's behavior to discover unknown problems.
\end{definition}

Proactivity in maintaining ML pipelines is especially challenging because the user needs to be able to anticipate issues with both the data and the models. Some ML-specific proactive queries could be:

\begin{itemize}
    \item Does the pipeline require continuous training \cite{continuous} to prevent the performance from degrading in the near future?
    \item Was there data leakage \cite{leakage}, or are there assumptions made in training that may not hold in live inference?
\end{itemize}

Table \ref{tab:userreqs} describes our requirements for ML pipeline observability, categorized by retrospectivity, proactivity, and user friendliness. \shreya{should we include slicing \& active learning / labeling?}

\begin{table}[h]
    \centering
    \begin{tabularx}{\columnwidth}{|X|X|X|}
    \hline
      \bf ID  &  \bf Requirement & \bf Category \\
     \hline
      RET1  & End-to-end tracing for any prediction & Retrospective \\
      \hline
      RET2 & Suggestions for where to begin debugging given faulty outputs & Retrospective \\
      \hline
      PRO1 & Alerts when performance might be going down & Proactive \\
      \hline 
      PRO2 & Suggestions for when to retrain models or rerun stages of the pipeline & Proactive \\
      \hline
      PRO3 & Test suite runner to assert expectations for data and transforms & Proactive \\
      \hline
      USE1 & Easy to incorporate into a tech stack with multiple different languages and tools & Usability \\
      \hline
      USE2 & Easy for people of different roles (engineer, data scientist, etc) to use and collaborate on & Usability \\
      \hline 
      USE3 & Minimal new things to learn & Usability \\
      \hline
    \end{tabularx}
    \caption{User requirements for an ML pipeline observability tool}
    \label{tab:userreqs}
\end{table}

\subsection{Challenges}
\label{sec:challenges}

We analyze challenges at the storage, execution, querying, and interface layers for each of the requirements. Different stages of the pipeline will require different logging patterns. For example, an ETL-heavy workload to create features may require logging the result of every \texttt{df.count()} function after every join to make sure rows are not dropped, although the user may not explicitly make the function call in the code. A training workload may be parallelized, for example, when doing a hyperparameter search, but require logs from each run to be sorted and concatenated. A challenge in defining a logging abstraction is to be flexible for different use cases but rigid enough to automatically log relevant information without the user explicitly coding up each idea they want to log.

\subsubsection{Storage layer}

\topic{Record-level data flow} Inputs and outputs from a coarse-grained data flow tracking mechanism could be simple to store as a graph, but the number of nodes and edges grows significantly as the logging gets more fine-grained. For example, a model could have millions of dependent data points, each of which come from many different transformations of upstream data sources. 

\topic{Time Travel} In order to maintain "snapshots" of the pipeline over time, we will need to store versions of all the data and artifacts. In many cases, users can have retrospective queries that date back months or years. Simply deleting data when a defined time period has passed is not an option for companies under stringent legal requirements such as banks. However, not all data or artifacts get used in production pipelines -- for example, a training job could ingest a table with thousands of columns and only use a handful of columns. Space could be saved using fine-grained provenance tracking coupled with triggers to delete unused data from the past.

\topic{Model storage} Data versioning has a rich history \shreya{cite here}, but in an environment where each data scientist might train thousands of models for an experiment, it is imperative to store models efficiently. Raw diffs between model versions can be as large as the models themselves. Some work in this area explores read-only optimized parameter archival storage systems for deep learning models \cite{modelhub}. In practice, many organizations have their own model abstractions that wrap around existing modeling libraries and store metadata such as training dataframes in addition to the model binaries. Efficiently storing large amounts of serialized files without having much insight into the contents of these files can be challenging.

\topic{Materializing data points for monitoring} One could imagine a user toggling a "monitor" switch for any inputs or outputs for any stage in the pipeline. We would need to run statistical aggregations and tests on both sliding windows of recent inputs and outputs to various stages of the pipeline as well as training sets for models. Furthermore, depending on input and output dimensionalities, number of metric types to compute, and how frequently the monitoring job would run, the number of metric values to store might be very high.

\topic{Privacy} The storage layer should support various levels of permissioning for every piece of data, model, or artifact. It becomes easy to accidentally mix data in the ML setting, when models may be trained, checkpointed, and fine-tuned on different data. An basic observability tool should store access patterns and compute whether users are properly adhering to permissions. An ideal observability tool should safeguard users from accidentally mixing data and violating permissions, as well as easily flag any dependent models or artifacts when some data is deleted from the system.

\shreya{static vs dynamic data}

\subsubsection{Execution layer}

% We can use the logging mechanism to track end-to-end data flow: in addition to logging metadata at runtime for any stage in a pipeline, we can log the inputs and outputs at each stage, such as features or models. 

\topic{Language and framework agnostic} The logging mechanism should be language and framework agnostic (USE1), which limits how fine-grained logs can be. Stages of the pipeline may be run on a distributed system (i.e. Spark cluster), which poses an additional challenge. Running a tracer at the line execution level for each stage in the pipeline can have an expensive overhead, so a fast message-passing mechanism is necessary. \shreya{maybe include slowdown for python decorator} Finally, logs may need to be sent to a remote server so multiple people can interact with them.

\topic{Efficient and effective metric computation} First, determining whether empirically observed groups of data points are significantly different is an unsolved algorithmic problem in the case where the shape of the distribution is unknown. Computing simple metrics like the mean and median is a good start but can fail when skew and kurtosis changes. Computing well-known distance metrics like Jensen-Shannon divergence or the Kolmogorov-Smirnov test can be expensive and produce too many false positive alerts. Second, even when the user knows what distance metrics they want to monitor, computing these metrics can be computationally expensive and redundant when two features represent similar concepts. We can leverage approximate query processing and batch queries to compute these distance metrics, but it will be challenging to choose what metrics to monitor for the user and eliminate redundancy in order to alert the user when pipeline performance may be decreasing (PRO1).

\topic{Testing} ML pipeline creators may want to perform certain tests on data-related triggers. For example, maybe the user will want to recompute outlier bounds for preprocessing when a handful of rows are added to a table. The DBMS field has a rich history of constraints and triggers that can be leveraged to perform basic monitoring and testing, keep stages of the ML pipeline from becoming stale (PRO2), and test for data integrity (PRO3).

\topic{Debugging guidance} Computing a trace for an output (RET1) is easy to make fast, but consolidating traces for a large group of outputs and identifying shared "nodes," or similar runs of a stage (RET2), can be challenging to do quickly. A naive pathfinding algorithm and ordering of nodes by frequency scales with how fine-grained the logged stages are and how many outputs a user wants to trace.

\subsubsection{Query layer}

\topic{SQL-based} Users should not have to learn another query language to ask questions about their pipeline (USE3). We can define a small set of commands to represent different functionalities such as computing a trace or inspecting the history of a stage in the pipeline. If the user wants to ask more fine-grained questions about the logs, they should be able to use an existing query language like SQL.

\topic{Forwards, backwards, and slice-based querying} In line with RET2 and PRO1, a user may want to query many metrics for many different data points. They also may want to query "forwards" and "backwards" through traces of data points -- for example identifying all the outputs influenced by a single model binary or all the artifacts involved in producing a single output. Finally, the user may want to perform slice-based querying to analyze pipeline performance on a subpopulation of their choice. Satisfying all of these types of queries in a single tool may be challenging.

\subsubsection{Interface layer}

\topic{Flexible logging abstraction} Existing software metric logging languages such as PromQL for Prometheus are not intuitive for writing data science metrics, which contributes to the lack of monitoring in many ML pipelines today. Different stages of a pipeline will have different logging needs and patterns. Additionally, at every stage, users will want to run certain tests and monitor specific metrics. Defining these abstractions may be challenging.

\topic{Presenting lots of information to many different stakeholders} The querying interface should be intuitive for both data scientists and engineers, which will require both a graphical UI and a CLI. A challenge is to distill information about thousands, or even millions, of metrics and outputs into a presentation that will not overwhelm the user. Multiple people at an organization should be able to access all the logs and run queries, so the interface needs an option to be hosted on a remote server (USE2). 

\topic{Integration with existing tools} The interface should present ways for the user to drill deeper into an issue or logs for a stage, such as links to existing tools used such as \texttt{kubectl logs} or Weights \& Biases reports. The user may want to define these integrations, since every user will likely have a different technical stack.

\end{full}

\section{Research Challenges}
\label{sec:researchprobs}
%!TEX root = main.tex

% \mr{While our roadmap involves incorporating existing, well-studied debugging techniques highlighted in Section~\ref{sec:related}, this paper focuses on unaddressed research challenges in ML observability related to detecting, diagnosing, and reacting to bugs. In detecting bugs, we focus on the problem of approximating live ML performance without ground-truth labels. In diagnosing bugs, we focus on suggesting constraints on input and output columns and tracking slow distribution shift. In reacting to bugs, we discuss a data-centric approach to maintaining a dynamic training set that yields high performance on live data.}

\mr{Before we describe our research challenges, we first introduce key definitions and an example ML pipeline (Figure~\ref{fig:tradpipeline})\techreport{ to ground our discussion}.} 

\subsection{ML Pipeline Preliminaries}

\subsubsection{Definitions}

\techreport{Here, we define several terms used throughout this paper.} 
An ML pipeline involves multiple data processing components, leading to one or more ML models that provide \emph{predictions} for a specific \emph{task}. A \emph{metric} is a measure of success for an ML pipeline, such as prediction accuracy. A \emph{tuple} is an individual feature vector used to generate predictions. A \emph{live} prediction is a prediction made after deployment, as opposed to predictions made during training. The consumers of predictions provide \emph{feedback}, or some data that indicates the quality of a prediction (e.g., item selection for recommendations, correctness for binary classification). \emph{Labels}, or "ground-truth" for predictions, are derived from feedback. Finally, we refer to groups of tuples, defined on conjunctions of predicates on features, as \emph{buckets}.\techreport{ A \emph{bucketing strategy} refers to how tuples are assigned to buckets.}

\subsubsection{Example ML Pipeline} 
Using data from the New York City Taxi and Limousine Coalition~\cite{nyc-taxi}, our ML task is to predict whether a rider will give their driver a tip > 20\% of the fare. Predictions are probabilities (i.e., floats between 0 and 1). Each tuple in the dataset (Yellow Trips) represents a single ride, with 17 attributes.

Our ML pipeline includes five components, as described by the rectangular boxes in Figure~\ref{fig:tradpipeline}. We have two sub-pipelines---training and inference---that share the cleaning and feature generation components. The pipeline includes one model, an \texttt{sklearn} random forest classifier.
The ML pipeline is evaluated on accuracy, or the fraction of correct predictions\techreport{,
when the prediction is rounded to the nearest integer}.

\begin{figure}
\papertext{\vspace{-10pt}}
    \centering
    \includegraphics[width=\linewidth]{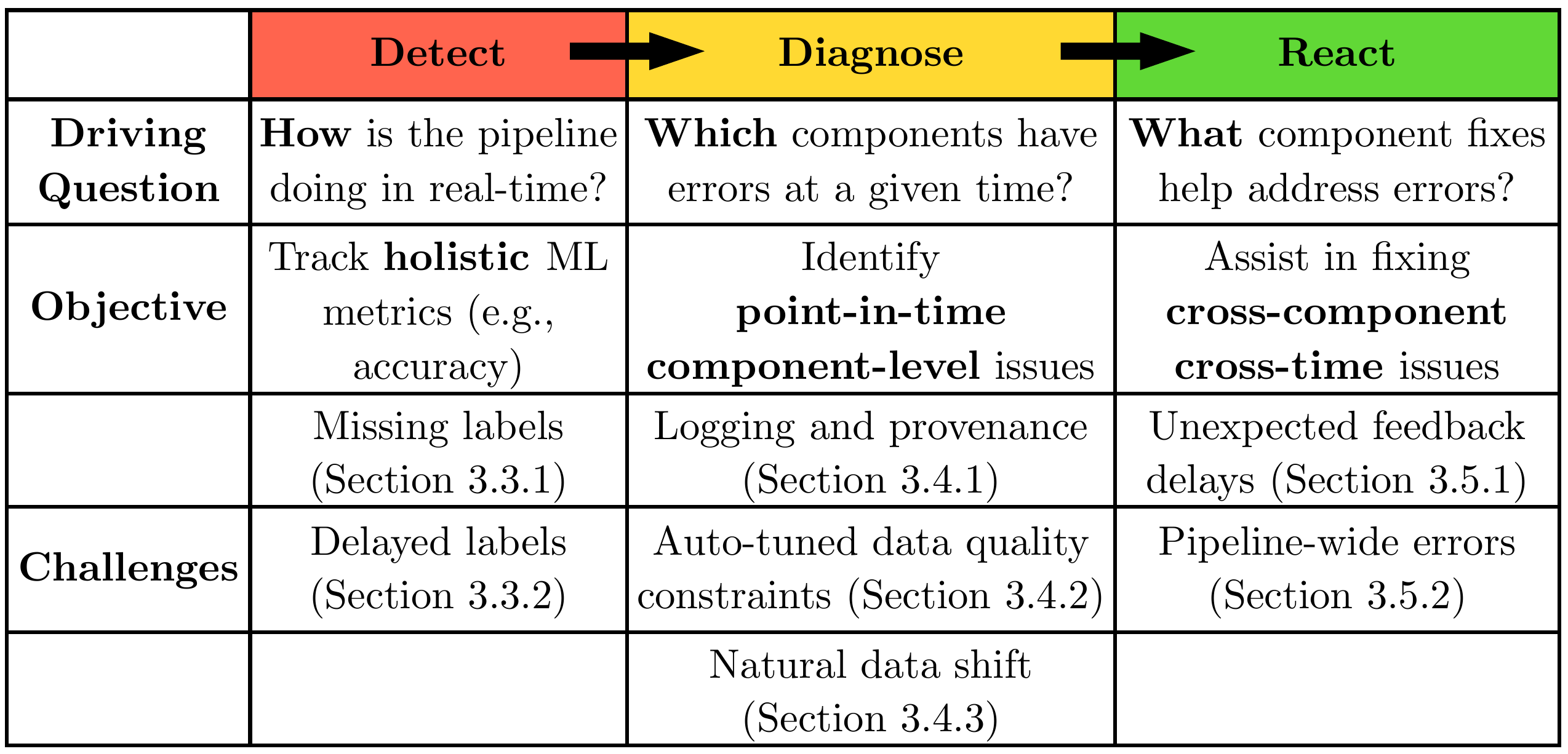}
    \vspace{-20pt}
    \caption{\mr{Breakdown of research challenges}}
    \label{fig:roadmap}
\papertext{\vspace{-15pt}}
\end{figure}

\subsubsection{Formalizing Distribution Shift}
\label{sec:distshiftformal}
\mr{In deployment settings, real-time accuracy is often difficult to measure due to feedback delays, so practitioners monitor changes, or shifts, in distributions of features and predictions}. ML practitioners have introduced any number of types of shifts, such as \mr{concept, data, covariate, label, subpopulation, prior probability, and low-data shifts, among others}---and these definitions often conflict~\cite{sugiyama, unifyingview, bbse, breeds, Wiles2021AFA}. If $Y$ is the label space and $X$ is the feature or covariate space (e.g., location of ride, number of passengers), we note that all of the aforementioned shift definitions boil down to {\em at least one} of the two shift scenarios:

\vspace{1pt}
\noindent {\bf Concept shift}: $P(Y | X)$ changes; $P(Y)$ changes but $P(X)$ doesn't \\
\noindent {\bf Covariate shift}: $P(X)$ and $P(Y)$ change but $P(Y | X)$ doesn't
\vspace{1pt}

\noindent A concrete example of concept shift is a recession: riders tip less, changing $P(Y)$ but not $P(X)$. A concrete example of covariate shift is New Year's Eve: the number of taxi rides will be relatively higher near Times Square in New York, changing $P(X)$ and $P(Y)$ as a result, even though the nature of a taxi ride that results in a high tip does not change, i.e., $P(Y | X)$. \rtwo{There's ML literature on learning under these natural shifts~\cite{gamasurvey, Lu2019LearningUC}; however, our research challenges focus on the \emph{unexpected} combinations of shifts that arise in production.}

The rationale for tracking $P(Y)$ and $P(X)$ over time \rtwo{in production pipelines} is that significant changes in these values can indicate \rtwo{new data quality or engineering bugs that need to be fixed.} 
% \strikeoutwhen{when and how to retrain models. For example, concept shift might imply a retrain over fresh data, whereas covariate shift might imply upsampling of certain populations in the data.} \agp{possibly delete? previous statement} \shreya{deteted} 
However, methods to flag changes in distributions, as mentioned in Section~\ref{sec:detectingdatashift}, cause too many false positive alerts. For example, practitioners compute the K-S test statistic between training and live tuples for \emph{each feature} to approximate how $P(X)$ has changed, \mr{often yielding thousands of measures, which can be confusing to navigate}.\techreport{  For instance, what would a user do with an alert saying a handful of their thousand features' K-S test statistics are now statistically significant? Does this alert really impact ML accuracy?} Additionally, \mr{on large datasets, $p$-values can go to zero even without actual significance~\cite{pvalues}, further exacerbating alert fatigue}.

% \shreya{Practitioners measure changes in $P(Y | X)$ and $P(X)$ by computing distances (e.g., KL divergences, K-S test statistics) between tuples in the training set and a window of live inference points. For example, one can compute the K-S test statistic between training predictions and live predictions to approximate how $P(Y | X)$ has changed. One can compute the K-S test statistic between training and live tuples for each feature to approximate how $P(X)$ has changed, but this procedure yields as many measures as there are features --- which can be thousands in practice.}

% \mr{To improve precision on real-time model performance alerts, we break down identifying errors into \emph{detection} and \emph{diagnosis} categories. We use model-specific metrics that require labels (e.g., accuracy) to detect when there are errors and finer-grained distance metrics that do not require labels (e.g., K-S test statistic between a feature's distribution in the training set and its live distribution at inference time) to explain where the bug lies in the pipeline (e.g., a feature's distribution shifted).
% In the following subsections, we discuss measuring model performance in the absence of labels (Section~\ref{sec:coarsegrained}) and diagnosing performance drops through column constraints and tracking distribution shift on provenance snapshots (Section~\ref{sec:finegrained}).
% Finally, we describe data-centric approaches to aid users to fix these issues (Section~\ref{sec:reacting}).}

\subsection{Research Roadmap}
\label{sec:roadmap}

\mr{As shown in \Cref{fig:roadmap}, we employ a three-pronged framework of detecting (Section~\ref{sec:coarsegrained}), diagnosing (Section~\ref{sec:finegrained}), and reacting (Section~\ref{sec:reacting}) to bugs in ML pipelines after deployment:
}

\smallskip
\noindent \mr{{\bf Detection.} This prong answers the 
question {\em how is the deployed ML pipeline doing in real-time}, 
with a focus on performance 
measures such as accuracy. 
There are two challenges 
in estimating performance. First, the lack of labels, 
which happens
soon after deployment (Section~\ref{sec:nofeedback}),
and second, labels are available but arbitrarily delayed (Section~\ref{sec:partialfeedback}). 
In the latter case, estimating real-time performance requires a
join across the out-of-sync label and prediction streams, 
which is difficult 
at scale.}

\smallskip
\noindent \mr{{\bf Diagnosis.} Given a drop in performance, 
diagnosis answers the question {\em which components
of the pipeline are potential sources of errors},
with a focus on a single component and a single point in time.
To make sense of errors in individual components,
we need to log 
intermediate inputs/outputs and provenance (Section~\ref{sec:provenance}).
With this logging in place, 
we should automatically specify and tune data validation constraints to address spectrum of pipeline bugs,
from hard and soft constraint violations (Section~\ref{sec:dataconstraints}), to data shift (Section~\ref{sec:distshift}).} 

\smallskip
\noindent \mr{{\bf Reaction.} With errors in individual components identified, 
reaction answers the question {\em what fixes to the pipeline
can help address errors}, across components and time.
To help fix the pipeline, we need to both make sure that any 
sources of label lag are addressed (Section~\ref{sec:react-label})---to ensure 
that we have better estimates of performance measures,
and that the cross-component and cross-time issues are addressed (Section~\ref{sec:react-component}).
}

\subsection{Detecting ML Performance Issues}
\label{sec:coarsegrained}

\rtwo{Post-deployment, the starting point for identifying issues
is monitoring drops in ML metrics such as accuracy. 
This becomes challenging when labels 
or predictions are delayed or absent. 
Moreover, delays may not be uniform across
buckets
(e.g., a power outage in East Village 
might prevent taxicab meter information 
from being uploaded). 
As shown in Figure~\ref{fig:tradpipeline}, 
predictions and feedback arrive 
at different timestamps and 
are joined on some identifier. 
At every timestamp, ML pipelines can move 
between three feedback scenarios: 
no feedback, partial feedback, and full feedback. 
There are two key challenges:
first, the lack of labels (impacting the partial and no feedback settings),
and second, arbitrary label delays (impacting the partial and full feedback settings).
We discuss both of these challenges in turn.
\papertext{We focus on 
cumulative accuracy, which is the easier case;
there are additional challenges in computing accuracy on sliding windows,
which we cover in our technical report~\cite{shankar2021observability}.}
}

\techreport{Before we discuss the challenges,
when estimating real-time accuracy, 
there are at least three variants of interest:
(a) cumulative accuracy for all predictions
made until now;
(b) accuracy for predictions made in
the last time window $t$;
(c) accuracy for the last $k$ predictions.
The last two variants provide accuracies over a sliding window.
The cumulative setting is not just relevant 
when we are evaluating accuracy from $t=0$;
it is also useful when we 
reset the clock regularly, 
e.g., accuracy
on a per-day basis.}

% \rtwo{There are a handful of detection challenges, such as large-scale joins of prediction and feedback (i.e., labels), efficiently tracking custom distance metrics between current and historical model score distributions, and computing pipeline metrics with low latency, that can be solved with straightforward adaptations of prior work \shreya{TODO: cite}. However, a major pain point faced almost universally ---  many false positive ML performance drop alerts --- is unaddressed. We propose approximating top-line ML metrics (e.g., accuracy) and monitoring changes in real-time; the biggest challenge here is that labels on predictions 
% can be delayed. \shreya{TODO: cite} Moreover, delays may not be uniform across 
% different buckets 
% (e.g., a power outage in East Village 
% might prevent taxicab meter information 
% from being uploaded) and 
% can be exacerbated in situations where 
% manual labeling is required. 
% As shown in Figure~\ref{fig:tradpipeline}, 
% predictions and feedback arrive 
% at different timestamps and 
% are joined on some identifier. 
% At every timestamp, ML pipelines can move 
% between three feedback scenarios: 
% no feedback, partial feedback, and full feedback, impacting real-time accuracy scores. We discuss each of the feedback scenarios in turn.}

\subsubsection{The Lack of Labels}
\label{sec:nofeedback}

\mr{After deployment, it is common to 
either have no labels or only a subset of predictions labeled.
These labels may arrive in batches at a later date, 
possibly after human review, motivating us to 
still find ways to estimate real-time performance without them.} To estimate cumulative accuracy, 
we may use importance weighting 
(IW) techniques~\cite{sugiyama}. 
We can identify buckets 
\mr{based on input feature combinations, 
determine the training accuracy} 
for each bucket, and weight these accuracies 
based on the number of points in each bucket 
in the live (post-deployment, unlabeled) data. 
Consider neighborhood 
as a bucketing strategy: if the 
training set had FiDi and Midtown accuracies 
of 80\% and 50\% respectively and we have 
100 FiDi and 500 Midtown live predictions, 
we can estimate an accuracy of $0.8 \times 100 + 0.5 \times 500 = 55\%$.

\begin{figure}
\papertext{\vspace{-8pt}}
    \centering
    \includegraphics[width=\linewidth]{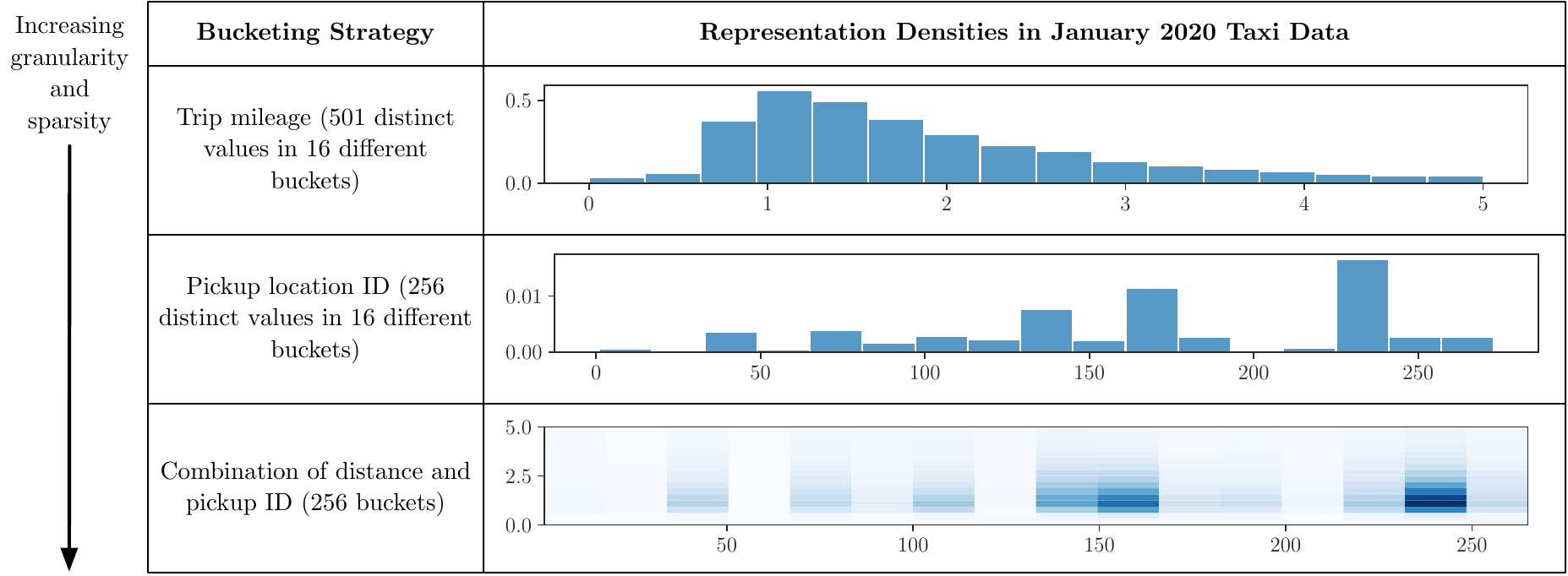}
    \vspace{-15pt}
    \caption{Bucketing strategies, normalized to show bucket density. As buckets become more finer-grained, they also become sparse.}
    \label{fig:bucketings}
    \papertext{\vspace{-20pt}}
\end{figure}

% \begin{figure}
%     \centering
%     \begin{subfigure}[b]{0.49\linewidth}
%         \includegraphics[width=\linewidth]{figs/feb1.pdf}
%         \caption{Difference between pickup locations in month of January 2020 and the week of February 1, 2020.}
%     \end{subfigure}%
%     \hfill
%     \begin{subfigure}[b]{0.49\linewidth}
%         \includegraphics[width=\linewidth]{figs/mar28.pdf}
%         \caption{Difference between pickup locations in month of January 2020 and the week of April 18, 2020.}
%     \end{subfigure}
%     \caption{Example of bucketing strategy based on pickup location. The distributions may diverge more as time goes on, potentially impacting the correctness of the accuracy estimate in the no-feedback case.}
%      \label{fig:locationhists}
% \end{figure}

\mr{There are multiple competing objectives
in determining which bucketing strategy would lead to best estimates
of accuracy,
among the $O(n!)$ possible bucketing schemes, 
where $n$ is the number of features, i.e., any subset of input features.
}
Figure~\ref{fig:bucketings} illustrates three bucketing schemes. 
The first couple have representation 
in each bucket, which gives us some confidence in per-bucket accuracy. 
However, the last one \mr{has some empty buckets}---so if a live tuple 
were to be assigned to such a bucket, we would not have an accuracy estimate for it.
Overall, finer-grained bucketing schemes may capture
patterns not found in coarse-grained ones
but could also 
be more sparse, which can impact the 
correctness of our accuracy estimates.
\mr{Moreover, finer-grained bucketing schemes would occupy
more space than coarse-grained ones.
Beyond {\em (i) Space} and {\em (ii) Sparsity}, 
there are other objectives we need to consider.
{\em (iii) Variance:} the buckets should 
have high variance in training accuracies, leading to more useful 
estimates.
{\em (iv) Predictiveness:} the training accuracy for each bucket
should be predictive of actual accuracies for live data
in that bucket.
Balancing these objectives is non-trivial. 
We may take inspiration from stratified sampling~\cite{parsons2014stratified}
in Approximate Query Processing (AQP)~\cite{agarwal2013blinkdb,acharya1999aqua},
and also ensemble schemes~\cite{sanjayaggaqp}.
}

\techreport{Extending this technique to the sliding window accuracy setting if
we are using a fixed offline bucketing may be straightforward.
Per bucket, we can apply ideas from prior work 
in streaming algorithms to update 
counts for the last $n$ tuples~\cite{datar2002maintaining}; 
similar techniques may also apply for the sliding window defined by time.}

Finally, we may gain additional benefits from changing
the bucketing strategy in response to live data.
\mr{To do so, we must}
efficiently identify buckets in high-dimensional, 
changing data streams with a reference dataset 
in mind (i.e., the training set). 
A starting point could 
be to extend streaming clustering algorithms 
that are explicitly robust to changing 
data distributions~\cite{Mousavi2015DataSC}: 
in addition to the live data, 
we could feed the training set to such a clustering algorithm.

\subsubsection{Label Delays}
\label{sec:partialfeedback}

\mr{
Labels are often delayed in arbitrary ways.
So, estimating accuracy, which requires a join between
prediction and label/feedback streams,
is challenging to do at scale, 
since keeping both streams in memory is impossible.
One option is to uniformly subsample both streams,
since the usual problems with sampling over joins~\cite{chaudhuri,kandula, joinstheoreticalguide} don't apply
when each prediction tuple joins with precisely one feedback tuple.\techreport{(It is well-known that uniformly subsampling streams before a join can yield quadratically fewer tuples in the result~\cite{chaudhuri}. 
Unlike the standard join setting, here, each prediction tuple
joins precisely with a single feedback tuple, 
meaning that the challenges of 
quadratically fewer samples 
with AQP over joins do not 
apply~\cite{chaudhuri,kandula, joinstheoreticalguide}.)}
} \mr{Since we do not know the size of the streams, 
one can apply}
reservoir sampling~\cite{aggarwal2006biased} on both streams
using a shared hash function on the common identifier.
However, this approach is wasteful, since
once the pair of prediction and feedback tuples are received,
they no longer both need to be in memory.  
Moreover, the quality of the estimate degrades over time 
since we are maintaining a fixed size sample
over growing streams.
Ideally we want to maintain both a
reservoir (for prediction tuples without feedback) 
plus partial aggregates (for prediction tuples with feedback).
Joined tuples can make way for new slots in the reservoir. 
However, doing so while respecting 
the reservoir sampling guarantee
of each having the same probability of being
sampled, is non-trivial.
For example, the sudden arrival of
a number of feedback tuples
can cause multiple slots in the reservoir to become
vacant, leading to an increasing probability for
the next prediction tuple to be included in the reservoir.

\techreport{Extending this reservoir sampling 
approach to (b) and (c) is also challenging. 
We can leverage prior work on 
reservoir sampling over 
windows~\cite{babcock2001sampling,aggarwal2006biased,gemulla2008sampling,bailis2017macrobase},
where we can evict old tuples from the reservoir
when they expire~\cite{babcock2001sampling}, or update the probabilities to
favor newer tuples more, using
an exponential decay weighting~\cite{bailis2017macrobase}. 
As before, we will want to modify these techniques to be less 
wasteful of memory, while also ensuring that they are unbiased.}

\techreport{Finally, we'd want to combine our techniques for dealing with label 
delays with that for dealing with missing labels (Section~\ref{sec:nofeedback}).
This is because 
many ML pipelines will be in the partial-feedback setting: 
often, live data is only partially labeled or arrives 
on a specific schedule. Or, some upstream 
data collection issues might 
delay feedback (e.g., there’s 
a cell tower outage in a region of Tribeca, 
causing payment meter data to be delayed). 
Here, aggregating the full-feedback 
and no-feedback estimates, weighted by the count of tuples in each case, 
may produce a reasonable real-time accuracy estimate.}

% \subsubsection{Monitoring Business Metrics}
% \label{sec:businessmetrics}

\rthree{Beyond labels, another way to approximate ML pipeline performance is to directly monitor changes in an important business metric (e.g., user satisfaction, click-through-rate, revenue). Sometimes the ML metric (e.g., model accuracy) does not align with a business metric (e.g., user satisfaction, revenue), requiring users to rethink the ML objective or discard the model altogther. An engineering challenge is to provide integrations with other system components that are not directly part of the ML pipeline---e.g.,  sales tools that record metrics like daily active users. To help users understand the effectiveness of their ML models, we can show correlations between ML metrics and business metrics over time.}

\subsection{Diagnosing ML Performance Issues}
\label{sec:finegrained}

%\agp{Mention refs: All three prongs are currently pain points for practitioners and require a lot of manual investigation, motivating opportunities for automation~\cite{metaobstalk, lyftarticle}. Circuitous and unmanageable ways of detecting bugs due to a lack of ground-truth labels yield false positives (as discussed in Section~\ref{sec:distshiftformal})}

\rtwo{After detecting an ML performance drop, 
we next need to diagnose it by identifying 
\emph{which} components have bugs at that time.
We focus on bugs 
that arise \emph{after} a pipeline deployment 
(i.e., related to a mismatch 
in data between training and serving).
There is a spectrum of data-centric ML production bugs~\cite{metaobstalk, lyftarticle}:
{\em hard $\rightarrow$ soft $\rightarrow$ drift},
from most to least time-sensitive.
Hard errors, such as some data sources failing to ingest and 
resulting in missing feature values, need immediate attention. 
Soft errors, such as features having anomalous means, 
require more tedious manual investigation because 
of false positives: many columns 
can deviate significantly while only a few are 
responsible for pipeline performance drops. 
Hard and soft errors are both forms of {\em pipeline errors},
which are often addressed by engineering changes to pipeline components.
Finally, bugs can result from natural data drift,
causing model performance to slowly decrease; nevertheless, 
they require attention.
Unlike pipeline errors, data drift occurs naturally
as data evolves and models no longer
faithfully capture the underlying relationships. 
%They often require some form of retraining or domain adaptation.
We discuss the challenges addressing 
pipeline errors and drift errors next, after discussing 
a prerequisite: logging and provenance.
}

\subsubsection{Logging and Provenance}
 \label{sec:provenance}
  \mr{Logging at the component level helps us uncover whether 
  the output of a given pipeline component has an error. 
  Then, to trace errors across components, we additionally
  need provenance. \techreport{Without provenance, practitioners
  typically diagnose at the model feature 
  and prediction level and use "tribal knowledge" to 
  trace misbehaving features to potential upstream causes.}
  There is extensive work on logging and provenance, e.g., ~\cite{hellerstein2017ground,oinn2004taverna,cheney2009provenance,oppold2022provenance,freire2012making},
  and for ML and data science pipelines~\cite{chapman2020capturing,agrawal2019data,namaki2020vamsa,hellerstein2017ground,chirigati2016reprozip, provdb,mlflow}.
  Some approaches require 
  using a specific end-to-end ML framework~\cite{agrawal2019data} or logging API~\cite{chapman2020capturing,hellerstein2017ground, provdb, mlflow}.
  Others instrument the AST or bytecode to capture lineage~\cite{mlinspect,guo2012burrito,chirigati2016reprozip}
  or employ error-prone static analysis~\cite{namaki2020vamsa}, 
  forcing users to remain in a particular language or ML framework.}
  \mr{We instead propose a simple bolt-on approach: users 
  annotate pipeline components (e.g., with decorators for Python) 
  with pointers to inputs and outputs (e.g., dataframe variables) 
  that automatically get logged to an observability store. 
  When users want to retrieve a trace, we 
  perform a depth-first-search through logs on the fly,
  to construct the provenance. 
  This approach offers a nice middle ground: not too onerous, while also letting
  users control how provenance is captured.}

\techreport{Unfortunately, logging raw inputs and outputs for each component in the ML pipeline
can quickly get expensive.
As an anecdote, the first author worked at a startup where 
the MLFlow~\cite{mlflow} logs would require a "purge" every few months.
To minimize log size, 
we can use the same approach as in the previous section and use
a reservoir sample for prediction tuples; a uniform sample may suffice for 
training tuples.
In addition, we can log histograms instead of full data streams; however, bins should change as data evolves over time. Research challenges lie in combining ideas from incrementally-maintained approximate histograms with ideas from adaptive histograms to produce evolving summaries of windows of data~\cite{fastincmaint}. 
Another insight is that users will only selectively query logged intermediates (e.g., inspect the head of a dataframe). For each component, we can learn from query patterns over time to inform what goes into logs, thereby reducing latency and storage footprints.} 

% Second, checking so many constraints
% when there are thousands of features
% can be quite expensive.
% We consider each issue in turn.
% When checking constraints, we may want to learn how to sequence
% the checking of constraints to reduce overall cost and quickly identify errors.
% This problem is reminiscent of work on adaptive query processing~\cite{avnur2000eddies,babu2004adaptive} by reordering predicates
% based on selectivity. We will need to adapt these techniques
% for the defined space of constraints---in our case, we may be able to identify
% the optimal constraint checking strategy offline. 

\subsubsection{Tracking Pipeline Errors via Auto-tuned Data Integrity Constraints}
\label{sec:dataconstraints}

\mr{ML pipeline errors are typically caught by 
data validation constraints~\cite{dataval, mlinspect, schelter, dde,dboost}.
For example, Schelter et al.~\cite{schelter} defines 
25 different types of single-column ML-specific constraints,
and two constraints
on column pairs, each requiring tediously
setting thresholds per column (or pair). The long-term maintenance of these constraints is also a headache: \papertext{this includes dealing with alert fatigue or missed constraint violations}\techreport{users spend months or years silencing hard constraints into soft constraints to avoid falsely rejecting predictions, converting soft constraints into hard constraints when they experience pipeline bugs}, and tuning thresholds for each constraint. Automating creation and maintenance for these constraints while preserving high precision (i.e., all violations correspond to bugs) and recall (i.e., all bugs are caught by violations) is an important challenge. Existing solutions suggest basic automatic constraints such as type checks and set membership for categorical columns; although they have decent precision, recall is low~\cite{tfx}. 
}

\techreport{Given provenance snapshots, we can execute constraint checks on intermediate inputs and outputs and log the results; when ML performance drops are flagged, users can then trace a prediction and inspect such results to determine which component(s) to address. The key question is, what constraint checks do we execute?} 
\mr{We propose auto-generating suites of input 
and output validation constraints and auto-tuning them over time to maximize precision and recall of violations. 
Constraints should be \emph{explainable} and 
thus more actionable for users, rather than seemingly random float-valued column bounds.} 
\techreport{If we have access to the user's model, we can narrow the search space by restricting constraints to columns involved in important features~\cite{Franois2006ThePT}; however, it is unrealistic to assume such access, especially for a bolt-on observability system.}
\rtwo{A solution idea is to learn an autoregressive model~\cite{cardestimation} that 
predicts the likelihood of a column's value 
for a tuple given values of the other columns; 
then we can aggregate likelihood scores for each 
column over sliding windows of post-deployment tuples. 
\techreport{Deep autoregressive models have demonstrated success 
in capturing rich multivariate distributions in challenging tasks, 
such as cardinality estimation~\cite{cardestimation}.} 
One approach is to fit an autoregressive function 
for each column to model its distribution at time $t$ 
given the joint distribution of other columns at time $t$, 
as well as tuples corresponding to times $< t$. 
However, this does not scale to more than a handful of columns, 
so we can use masking techniques to train a single model for all columns~\cite{made, bert, cardestimation}. 
Another challenge is that most off-the-shelf autoregressive models 
are trained explicitly to predict the next value or token, not to learn the \emph{distributions}, or density functions, of columns.
Thus, we need to discretize inputs and outputs 
so the model learns a distribution of buckets. 
A simple bucketization strategy can be derived from the CDF (e.g., quantiles), 
but this will fail as distributions change over time. 
Alternatively, recent progress in language modeling suggests 
a different bucketization strategy---numbers 
can be discretized into digits~\cite{Wallace2019DoNM}.}
\rtwo{To turn the density estimation model(s) into a suite of constraints, users can define a common threshold for columns, e.g., tuples should have an aggregated column likelihood score above 80\%. We can also fine-tune the threshold based on the size of the data (i.e., number of columns and tuples) to find a good trade-off between precision and recall of alerts. Further challenges include efficiently maintaining these autoregressive models and their data (e.g., fine-tuning, running across provenance snapshots).}

% \rtwo{While the data integrity constraints provide valuable guardrails to practitioners, two other challenges remain unaddressed. First, users may want suggestions for constraints to add, both to surface checks that they may have missed and to adapt to the changing nature of data. We can use anomalies flagged by data validation to indicate new constraints. Second, which the remainder of this section will focus on, is that data might drift over time, passing data integrity constraints.}

\subsubsection{Addressing Natural Data Drift} 
\label{sec:distshift}

\begin{figure}
\papertext{\vspace{-10pt}}
    \centering
    \includegraphics[width=\linewidth]{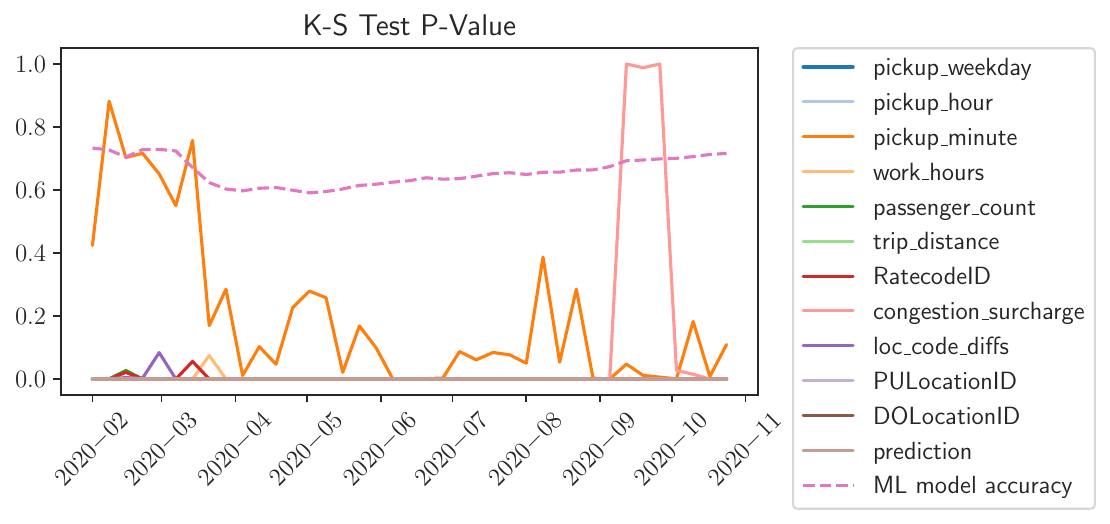}
    \vspace{-20pt}
    \caption{K-S test $p$-values for features and predictions. The training set (Jan 2020) is compared to \techreport{week-long }sliding windows of inference (Feb 2020--). $p$-values are "significant" ($<0.05$) throughout. \techreport{The ML model accuracy is depicted by the dashed line.}}
    \label{fig:pvalues}
\papertext{\vspace{-22pt}}
\end{figure}

Data integrity checks do not flag slower, longer-term distribution shift\techreport{, 
motivating the need to track how data distributions 
change over time}. For instance, a recession could cause riders to tip less across the population, changing $P(Y)$ but not $P(X)$. To approximately compute shifts in $P(X)$ and $P(Y)$, existing work proposes tracking metrics like KL divergence and KS tests~\cite{failingloudly} between sliding windows in live inference data and train datasets (i.e., for train-serve skew as described in Breck et al.~\cite{dataval}). There are two problems with this approach: (1) it requires the inference and training data to be kept in memory, and (2) it doesn't work well when there are many tuples---$p$-values go to zero even if shifts aren't significant enough to warrant a retrain, as discussed in Section~\ref{sec:distshiftformal} and shown in Figure~\ref{fig:pvalues}.

To solve (1), the memory issue, we can leverage a reservoir of live tuples (as in Section~\ref{sec:coarsegrained}), but it is impractical to keep the entire training set in memory. We can keep a materialized sample of the training set in-memory, but randomly sampling the training set might neglect important tuples, such as those from minority classes. As a solution, we can obtain a weighted random sample of the train set, where each tuple is weighted by its loss.

To solve (2), the $p$-value issue, we can draw inspiration 
from adversarial validation, a\techreport{ Kaggle community-originated} 
method to determine whether train and test sets 
are drawn from the same distribution~\cite{ellis_2021}. 
Adversarial validation trains a binary classifier $F$ to predict whether a tuple $d$ came 
from either the train or test dataset. 
If $F$ converges to $\sim$ 50\% AUC~\cite{ling2003auc}, 
then one can assume the datasets are 
similar~\cite{Pan2020AdversarialVA}. 
\mr{We can extend this to track shift}: 
we train $F$ to predict whether $d$ 
comes from the training sample or 
the reservoir sample of live data 
(as in Section~\ref{sec:coarsegrained}), and log the AUC. 
However, adapting this method to the streaming 
setting is computationally challenging because 
we would need to train a new classifier $F$ 
every time we log an AUC, 
and computing AUC requires multiple passes through the data.

\begin{figure}
    \centering
    \papertext{\vspace{-10pt}}
    \includegraphics[width=\linewidth]{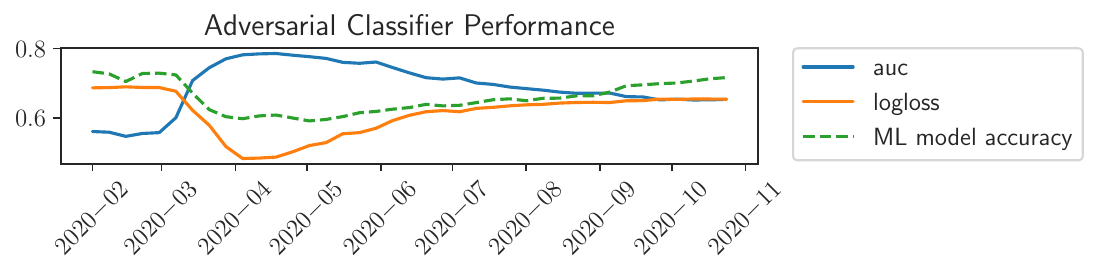}
    \vspace{-17pt}
    \caption{AUC and log loss from the adversarial classifier trained to separate a training data sample and reservoir sample of live tuples.\techreport{The ML model accuracy is depicted by the dashed line.}}
    \label{fig:advc}
    \papertext{\vspace{-15pt}}
\end{figure}

One insight is that users\techreport{ don't exactly care about the AUC,
they} only care about how the AUC changes over time, 
as an increasing AUC indicates that live data 
is diverging from training set data. 
As a proxy, we can log $F(d)$'s \emph{loss} over time, 
which can be computed in a single pass. 
To avoid frequently retraining $F(d)$ from scratch, 
every time we get a new tuple in the reservoir sample of live data, we can sample $d$ from the reservoir with $p=0.5$ and the training set with $p=0.5$; then, we can fine-tune $F(d)$ on $d$ with stochastic gradient descent. Here, the intuition is that decreases in loss are coupled with increases in AUC, as shown in Figure~\ref{fig:advc}. As loss decreases, it becomes easier to separate the training and live data, indicating distribution shift. The onset of distribution shift as flagged by the adversarial classifier aligns with the beginning of the ML model accuracy drop (late March 2020).
The features highly weighted in $F(d)$ are also
the ones most likely to be responsible for the shift,
further aiding diagnosis.

\subsection{Reacting to Bugs in ML Pipelines}
\label{sec:reacting}
\mr{Once bugs are isolated, they need to be fixed.
Slower distribution shift can be fixed by a retrain, 
but silent pipeline errors 
require immediate engineering attention. 
These pipeline errors 
require careful analysis across components 
and time: the challenge is to determine 
which pipeline errors that, upon fixing, 
will have the largest positive impact on ML accuracy.}
\mr{Unlike traditional data 
repair problems~\cite{dde, activeclean, holoclean, visclean, cleaningqamatching}, 
where the focus is on cleaning a snapshot of the data, 
here we want to point users to pipeline errors that, 
if addressed, 
can best improve future prediction quality. 
We discuss two such pipeline errors:
reacting to label feedback delays 
and repairing broken pipeline components.}

% \mr{In the case of an obvious engineering issue like a failed workflow run, a developer can make appropriate changes and relaunch the pipeline. Similarly, reacting to a violation of a hard data integrity constraint (e.g., too many nulls in a column\techreport{ because of a broken taxicab meter}) can be straightforward. But when soft data integrity constraints are violated or data drifts, users need slices of data to inspect and potentially retrain models on.}

 \subsubsection{Reacting to Feedback Delays}\label{sec:react-label}
\mr{Knowing how the distribution of 
feedback delays changes over time 
can uncover pipeline errors} 
and enable practitioners to quickly respond to them. 
Assuming the distribution of label delay 
is unknown and nonstationary 
(i.e., it may not be feasible to train a separate 
model to predict which predictions won’t have feedback), 
a challenge lies in identifying groups of 
tuples that have similar feedback delay times
to understand patterns. 
\mr{Most streaming clustering algorithms 
may not produce interpretable groups, i.e., those
described with only 
a few predicates}~\cite{Saisubramanian}. 
For debugging purposes, users may also care 
about how these clusters 
of delayed tuples change over time, or anomalies in delays\techreport{;
especially in the sliding window settings}.

\techreport{Consider the cumulative setting first.}
Overall, we want to pick predicate combinations that
"cover" all of the tuples that have severe label delays.
This is analogous to frequent itemsets~\cite{leskovec2020mining};
recent work has extended it to work in an approximate setting,
while optimizing for metrics like coverage~\cite{joglekar2017interactive}.
Unlike that setting, here, we cannot materialize a sample upfront
and operate on it; instead, we must operate on a stream
directly, and determine what predicate combinations may have high
coverage "on the fly".
For this, we can draw on incremental maintenance 
for frequent itemsets~\cite{incmaintfreq},
however this work focuses on 
updating itemsets given the addition of new tuples.
In our setting some prediction tuples that are missing feedback
may have their feedback arrive a bit later than expected.
Therefore, we will need to both add and remove tuples
and thereby update the counts of 
the current frequent itemsets during incremental maintenance.

\techreport{These challenges are exacerbated in the sliding window setting.
Here, we may be able to draw on work on 
streaming frequent itemsets~\cite{streamingfreqitemset,chang2003finding}.
For example, Chang et al.~\cite{chang2003finding} use time-weighting
to decay frequencies of itemsets over time unless
they were seen recently.
Doing this in the presence of feedback tuples appearing later 
in a delayed fashion is not straightforward.}

\subsubsection{Assisted Repair of Broken Components}\label{sec:react-component}

\rtwo{Besides feedback, 
there are two types of pipeline errors 
that cause performance drops: 
data staleness (no change) 
and corruption (unexpected change). 
A staleness example is if 
the pipeline to regenerate rider-related features 
(e.g., historical average tip) broke, 
forcing reads of old feature values. 
A corruption example is if an engineer 
changed geographical features to read 
from a better maps API, 
but the API returned distances in kilometers instead of miles.}

\rtwo{A key insight for both types of pipeline errors is that they are caused
by \emph{columns}, because columns tend to be 
outputs of pipeline logic (e.g., creating features)~\cite{dataval}. 
Quantitative data cleaning techniques 
from the statistics and database literature typically define units of data to be cleaned
as subsets of tuples, 
not columns~\cite{dde, anomalysurvey, quantdatacleaning}. 
Column-level changes are hard to catch---in our corruption example, 
a few anomalous trip distances isn't unusual, 
but all of them suddenly increasing is. 
We could leverage functional dependency (FD) 
discovery techniques to identify which columns most 
violate FDs~\cite{fdrepair, holoclean}. 
However, these are hard to apply in a noisy multivariate setting and 
consistently tune for production pipelines, 
especially without prior specifications from users.}

\rtwo{To assist repairing broken components, 
our research question is: 
what (component, column set) pairs best explain an ML performance drop? 
We can leverage the auto-tuned 
constraints 
from Section~\ref{sec:dataconstraints} to identify fuzzy changes 
in behavior at the column level. 
First, we determine individual column error scores 
for each component based on its historical behavior; 
then, we group columns by statistical correlations. 
Finally, we rank the (component, column set) error 
scores based on pipeline behavior by aggregating them across the dataflow graph. We discuss all three steps in turn.}

\rtwo{To measure column error, 
we can adapt statistical 
anomaly detection techniques to track how 
our auto-tuned data integrity constraints behave over time. 
\papertext{Concretely, since we have learned 
a representation of $\Pr[X]$ for any column $X$, 
we fix some scalar estimate of $\Pr[X]$ (e.g., $\mu(\Pr[X])$ or $p25(\Pr[X])$) 
and derive a time-based anomaly score
(e.g., how many standard deviations $\mu(\Pr[X])$ at a given time 
is away from its rolling, 7-day mean). 
Columns with an anomaly score $\geq 3$ may indicate errors. 
Note that this approach flags corruption bugs but not staleness bugs, 
which will have anomaly scores of 0. 
To formulate staleness as an anomaly detection problem, 
we derive the time-based anomaly score 
from a different scalar estimate of $\Pr[X]$, 
such as the variance of recent means of the column. 
Consequently, zero-valued historical variances and large-magnitude anomaly scores indicate staleness. We leave further implementation details to the tech report~\cite{shankar2021observability}.}} \techreport{
Concretely, suppose we have a column variable $X$, a representation of $\Pr[X]$ that determines our data integrity constraints, and seasonality window size $w$ (e.g, weekly). We derive an corruption error score $\zeta_X^{(t)}$ for column $X$ at time $t$ from the following z-score outlier formula:

% \begin{align*}
%     \omega_X^{(t)} &= \text{some aggregation of $\Pr[X]$, like } \mu\left(\Pr[X]^{(t)}\right) \\
%     \zeta_X^{(t)} &= \left| \frac{\omega_X^{(t)} - \mu\left({\omega_X^{(t - 1)}, \omega_X^{(t - 2)}, \hdots, \omega_X^{(t - w)}}\right)}{\sigma\left(\omega_X^{(t - 1)}, \omega_X^{(t - 2)}, \hdots, \omega_X^{(t - w)}\right)} \right| \label{eq:errorscore}\tag{1}
% \end{align*}

\begin{align*}
    \omega_X^{(t)} &= \text{some aggregation of $\Pr[X]$, like } \mu\left(\Pr[X]^{(t)}\right) \\
    \zeta_X^{(t)} &= \left| \frac{\omega_X^{(t)} - \mu\left(\bigcup_{i=1}^w \omega_X^{(t - i)} \right)}{\sigma\left(\bigcup_{i=1}^w \omega_X^{(t - i)} \right)} \right| \label{eq:errorscore}\tag{1}
\end{align*}
\smallskip

Intuitively, a larger $\zeta$ corresponds to more "anomalous" behavior (as per the definition of z-score). This approach is practical because column values do not need to be normally distributed for this method to work --- the z-score is being computed across a window of \emph{aggregations} of the density function $\Pr[X]$ over time, not across the column values themselves. To determine staleness error scores, we can also apply Equation~\eqref{eq:errorscore}: the challenge is to determine the aggregations of $\Pr[X]$ to set $\omega_X$ to.

Staleness is defined by $X^{(t)} = X^{(t - 1)} = X^{(t - 2)} = \hdots =  X^{(t - k)}$ for the $k$ time steps that a component has not refreshed. One issue in detecting staleness is that some components are designed to produce static features; for example, many user-related, dimensional features should not change. It would be unhelpful to tell a practitioner to repair such a component. Thus, we formulate staleness as an anomaly detection problem. Intuitively, the variance of a column statistic---e.g., mean---should not change if the column is stale. So, a simple solution is to set $\omega_X$ to the standard deviation of recent means:

\begin{align*}
{\omega_X^{(t)}}_{\text{STALENESS}} = \sigma\left(\bigcup_{i=0}^w \mu\left(\Pr[X]^{(t - i)}\right) \right)
\end{align*}
\smallskip

In this example, we chose the mean because it is unlikely, 
in large datasets, that column values change while leaving 
the mean unchanged. However, we can easily replace mean 
with a different statistic, e.g., $\Pr[X < \tau]$ 
for some threshold $\tau$, which could be a historical $p25$ or $p75$ value. We could also extend this to a multivariate setting to leverage multiple thresholds and more accurately compare distributions over time.} \rtwo{A practical challenge is that columns are highly correlated, but we can group co-erroneous columns based on correlation. 
One option is to use probabilistic 
graphical models~\cite{holoclean}, 
but a simpler, non-ML idea is to sample a 
covariance matrix to represent "distances" between features and apply spectral clustering.}

\rtwo{For the last step---ranking errors across all pipeline 
components---we track error across the dataflow graph. Each node represents a (column group, component) pair. 
Two nodes share an edge if their corresponding 
components share an edge in the ML pipeline graph. \techreport{Each node is labeled or weighted with its error score. Algorithm \ref{algo:errorflow} describes how error scores can be propagated across the pipeline---intuitively, intermediate nodes with small errors should deprioritize errors earlier in the pipeline, as our goal is to find nodes that cause high errors later in the pipeline. The algorithm outputs a list of sorted (column group, component) pairs that users can address, in order.}
\papertext{Each node is initially labeled with its error score $\zeta$. 
Intermediate nodes with small $\zeta$s should deprioritize 
errors earlier in the pipeline, as our goal is to find nodes that 
cause high $\zeta$'s later in the pipeline. For our ranking algorithm, we can conceptually traverse from the predictions node (i.e., the final component) 
backwards through all paths $p \in P$ in the graph; at every node $i$, we record $\phi_i = \max_{p \in P} \prod_j \zeta_j$ for all the nodes $j$ visited so far (including the current node $i$). Nodes with high $\phi$ require immediate attention.}}

% for each ordered path $p \in P$, at every node $j$, we record $\phi_j = \prod_k \zeta_k$ for all the nodes $k$ visited so far (including the current node $i$). Finally, for each node $i$ in the entire graph, we take the maximum recorded score: $\phi_i^{\text{MAX}} = \max_{p \in P} \phi_i$. Nodes with the highest $\phi^{\text{MAX}}$ require the most immediate user attention

\techreport{\begin{algorithm}
\SetAlgoLined
\KwIn{Final node $n_0$ (e.g., predictions node); \text{weight} and \text{neighbors} graph utility functions}
\KwOut{Ordered list of scores for (column group, component) nodes}
$E \gets \emptyset$\;
$q \gets \{(n_0, \text{weight}(n_0))\}$\;
\While{! q.isEmpty()}{
    $n_i, \phi_i \gets q.pop()$\;
    $\zeta_i \gets \text{weight}(n_i)$\;
    \ForEach{\text{neighbor} $n_j$ \textbf{in} \text{neighbors}($n_i$)}{ 
        \If{$n_j \in q$}{
            \If{$q[n_j] < \phi_i * \zeta_i$}{
                $q[n_j] \gets \phi_i * \zeta_i$\;
            }
        }
        \Else{
            $q \gets q \cup (n_j, \phi_i * \zeta_i)$\;
        }
    }
    $E \gets E \cup (n_i, \phi_i * \zeta_i)$\;
}
\Return{$\text{sort}(E)$}\;
\caption{{\sc ComponentRanking}: Tracks (column group, component) errors across a dataflow graph}
\label{algo:errorflow}
\end{algorithm}}

\techreport{\subsubsection{Case Study: Recommender Systems}
\label{sec:casestudy}

\mr{We illustrate the proposed diagnosis and reaction techniques with a short case study of real ML performance drops. Some details are modified and omitted due to NDA constraints. This case study involves an organization that is currently implementing the detection, diagnosis, and reaction framework to achieve ML observability.} 

\mr{The organization's flagship product is a mobile app with a large recommender system, which consists of approximately 20 separate ML prediction tasks (i.e., ML pipelines) that read from the same data sources. Each model takes in O(10,000) features, generated from different pipelines such as \texttt{user\_id\_features}, \texttt{content\_features}, and \texttt{engagement\_features}. Feedback delay is not the most important problem for this organization --- the main pain points for ML engineers include, when one of the models has a performance drop, identifying relevant broken columns and quickly determining the engineering issue that caused the performance drop. }

\mr{First, we describe an example of a staleness bug: the user ID features pipeline had not run successfully for weeks. This bug was detected by a drop in model accuracy. The ML pipelines simply read old \texttt{user\_id\_features} when making predictions, which performed poorly for new users that had recently signed up. The assisted repair tool analyzed soft integrity constraints on all the features and noticed high corruption scores for \texttt{user\_id\_features}, for $\omega_X = \Pr[X = NULL]$. This meant that the \texttt{user\_id\_features} output columns were becoming more null-valued, indicating more missing data. When slicing error scores by a popular user region (e.g., USA), the tool showed high staleness error scores. Thus, the organization was able to learn that the \texttt{user\_id\_features} pipeline had stopped running and make appropriate fixes.}

\mr{Second, we describe an example of a corruption bug: a recent release of the app did not have audio working. As a result, users could not engage with recommended content normally, and the organization detected a drop in model accuracy.  None of the hard integrity constraints (e.g., schema check, nonnegativity) were violated, and most distribution shift trackers (e.g., KS test) fired an alert for each column, which did not help with diagnosis or reaction. The assisted repair tool similarly found that most features had higher-than-usual corruption scores, but sound-related features (e.g., \texttt{audio\_on}, \texttt{num\_sound\_button\_toggles}) had the highest error scores. Thus, an on-call engineer was able to hypothesize that the audio might be broken and found a breaking PR in the changelog.}

\rthree{Finally, sometimes model performance boosts don't improve topline business metrics. For example, the model to predict the likelihood of a comment improves, but since users treat viewing, liking, and commenting differently, the average time spent viewing a recommendation can decrease. A separate engineering challenge for ML observability systems is to provide integrations with other system components that are not directly part of the ML pipeline --- e.g.,  sales tools to log metrics like daily active users --- such that users understand how much ML models impact business performance.}
}

\section{System}
\label{sec:system}

\techreport{
\begin{figure}[!t]
    \centering
    \includegraphics[width=\linewidth]{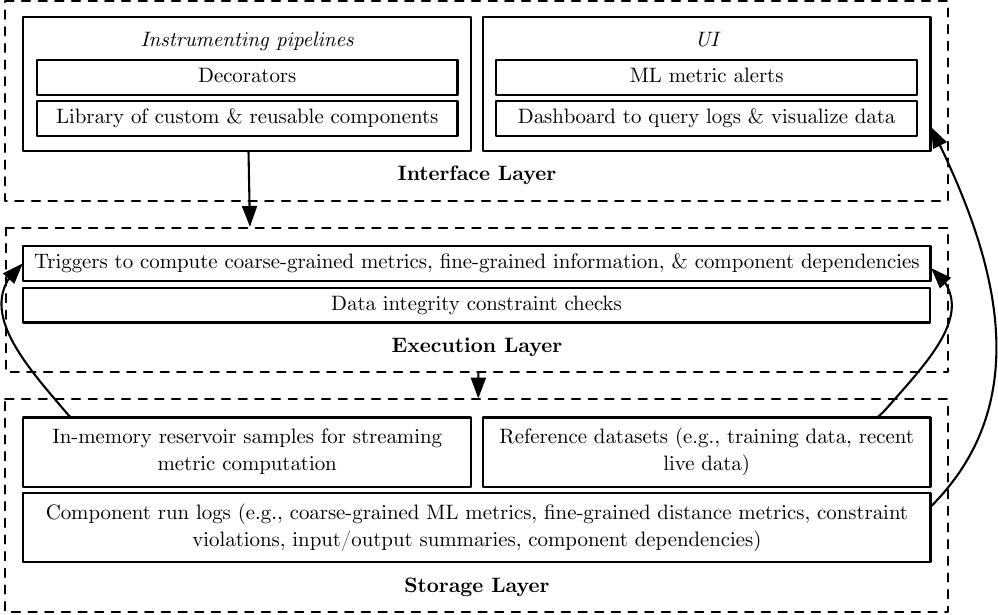}
\vspace{-15pt}
    \caption{Proposed ML observability system architecture.}
    \label{fig:architecture}
\end{figure}
}

A bolt-on ML observability system must be able to compute and store (1) history of and (2) interactions between components, requiring logging state at component runtime. Data and model integrity checks (e.g., expected number of nulls, model assertions~\cite{Kang2020ModelAF}) can be programmed as constraints. Metric computation (e.g., approximate accuracy) can run as triggers.

\topic{Interface Layer} Users should be able to view real-time pipeline performance (i.e., accuracy) and query fine-grained data summaries, traces for outputs, and other information in component logs. 
\techreport{Furthermore, in debugging low ML performance, users need interfaces to assess label delay, inspect data shifts and constraints, and react across components, motivating
techniques from visualization recommendation to highlight the most salient components~\cite{lee2021deconstructing, lee2021lux, wongsuphasawat2015voyager, wongsuphasawat2016towards}. }

\topic{Execution Layer} The execution layer, which wraps around a component, 
must be able to run trigger computation such as importance-weighting, executing and \mr{fine-tuning data quality constraints, drift detectors, time-based anomaly detection, and other methods described in \Cref{sec:researchprobs}. Additionally, the execution layer must identify component dependencies to track provenance for predictions}. 

\topic{Storage Layer} \techreport{As shown in \Cref{fig:architecture}, we}\papertext{We} must store at least three types of data: pointers to inputs and outputs, ML metrics monitored across consecutive runs of the same component (Section~\ref{sec:coarsegrained}), and logs capturing fine-grained state \mr{(provenance, data validation results, and drift detection as described in Section~\ref{sec:finegrained})} every time a component is run. Additionally, the system must keep samples of training sets and live inference tuples in-memory for the execution layer to use while computing fine-grained information (e.g., K-S test results, adversarial classifier weights).

\techreport{\subsection{\large \mltrace{} Abstractions}}

\papertext{\topic{\mltrace Abstractions}}\label{sec:mltrace} Our bolt-on ML observability system, \mltrace{}, will eventually have the following functionality:
(1) a library of functions that can support predefined computation before or after component runs for metric calculation or any relevant alerts, triggers, or constraints; (2) automatic logging of inputs, outputs, and metadata at the component run level; and 
(3) an interface for users to ask arbitrary post-hoc queries about their pipelines. Our current prototype has preliminary approaches for (2) and (3) and we are working on populating our library (1). We provide declarative, client-facing abstractions for users to specify components and the metrics and tests they would like to compute at every run of the component. \mr{The current prototype of \mltrace{} is publicly available on Github~\cite{mltracegit} and PyPI~\cite{mltracepypi}. 
\papertext{Additional details can be found in our technical report~\cite{shankar2021observability}.}}

\techreport{
\topic{Component} The \texttt{Component} abstraction represents a stage in a pipeline, similar to Kubeflow \cite{kubeflow} notation, and houses its static metadata, such as the name (primary key), description, owner, and any string-valued tags. The \texttt{Component} abstraction also includes \texttt{beforeRun} and \texttt{afterRun} methods for the user to define computation, or triggers, to be run before and after the component is run. These methods will primarily be used for testing and monitoring. \mltrace{} will have a library of common components that practitioners can use off-the-shelf, such as a \texttt{TrainingComponent} that might check for train-test leakage in its \texttt{beforeRun} method and verify there is no overfitting in the \texttt{afterRun} method. Additionally, users can create their own types of components if they want to have finer-grained control. 

\topic{ComponentRun} The \texttt{ComponentRun} (\texttt{CR} for short) abstraction represents dynamic metadata associated with a run or execution of a component. It includes the relevant \texttt{Component} name (foreign key), start timestamp of the run, end timestamp of the run, inputs, outputs, source code snapshot or git hash, extra notes, staleness indicator, and dependent \texttt{CR}s. Unlike other DAG-based tools, users do not need to explicitly define dependent components. \mltrace{} sets the dependencies at runtime based on the input values; for example, if a feature generation \texttt{CR} produced an output \texttt{features.csv} and an inference \texttt{CR} used \texttt{features.csv} as an input, \mltrace{} would add the feature generation \texttt{CR} as a dependency for the inference \texttt{CR}.

\topic{IOPointer} Inputs and outputs for a \texttt{ComponentRun} are represented by \texttt{IOPointer}s. In the current prototype, the \texttt{IOPointer} holds only a string identifier (e.g., \texttt{features.csv} or \texttt{model.joblib}) and its serialized raw data. We plan to make historical inputs and outputs available to users in \texttt{beforeRun} and \texttt{afterRun} triggers.

For \mltrace{} to be as light as possible, we only require users to interact with the \texttt{Component} abstraction. \texttt{CR}s and \texttt{IOPointer}s are created at component runtime via decorators on functions that represent component execution (e.g., the function that preprocesses data). 
}%techreport

\begin{full}
\section{Observability Query Patterns}
In this section, we discuss what metrics a user might monitor to assess ML pipeline health and how they might use \mltrace{} to ask questions about their pipelines for debugging purposes. 

\subsection{Service Level Agreements}

%\shreya{should we shorten this?}

Software engineering teams typically have \emph{Service Level Agreements} (SLAs) to ensure their applications provide business value. An example of a software SLA, or a contractual agreement to hit a target measure of service performance, could be 99.9\% uptime for a client-facing API-endpoint. Extending this concept, an example ML SLA could be 90\% recall for a pipeline that predicts taxi riders who will tip their drivers. It is imperative to monitor business-critical metrics, such as minimum accuracy or click-through rate.

However, several existing ML monitoring solutions in industry overemphasize tracking distributions of pipeline intermediates, such as features, and outputs. While this can be helpful for debugging purposes, these monitored values are not useful on their own. For instance, what would a user do if they received an alert that one of their thousand features' mean value dropped by 50\%? It is more fruitful to see if business-critical metrics have dropped when assessing whether components of pipelines are stale. Additionally, triggering alerts for all intermediates can contribute to alert "fatigue," rendering metrics useless in practice. To circumvent these problems, \mltrace{} houses intermediate aggregations in \texttt{ComponentRun} logs and focuses alert-triggering metrics on SLAs or other business-critical requirements. Users declare their ML-related metrics upfront in the \texttt{Component} abstraction, which are computed every time that component is run and trigger alerts when SLAs are not met. 

\begin{full}
\subsection{Service level terminology for ML}

\topic{Indicators} SLIs are quantifiable measures of service performance. Common examples of ML application SLIs are accuracy, precision, or recall over defined periods of time.

\topic{Objectives} SLOs are target values for SLIs. An example in ML could be an average accuracy of 95\%. Additional complexity arises for ML applications when tracking SLOs at a population subgroup level, which could be legally mandated (i.e., to ensure fairness).

\topic{Agreements} SLAs are contractual agreements to hit SLOs and primarily driven by business decisions. ML projects in production are often canceled because ML product teams do not meet their SLAs, resulting in little or even negative monetary return.

We posit that all user queries directly or indirectly address meeting their SLAs; thus our data model  treats SLIs as first-class citizens. Users declare their ML-related SLIs upfront in the \texttt{Component} abstraction, which are computed every time that component is run and trigger alerts when SLAs are not met. 
\end{full}

\subsection{Query Patterns}

We posit that all user queries directly or indirectly address meeting their SLAs. Next, we provide some examples of ML observability questions, describe how they fall into one of the four categories defined in Section~\ref{sec:userprobs}, and discuss how \mltrace{} can address them.
% \aditya{This is a weird use of "data model" (likewise
% in previous section);
% usually this refers to an abstraction capturing how
% the data is represented.}

\begin{example}
Why is there a large, sudden drop in accuracy?
\end{example}

The user traces outputs produced by the most recent inference component run. They perform a component run-level query: they look at the logs for each component run in the trace to see the results of the \texttt{beforeRun} and \texttt{afterRun} triggers. They find that the fraction of NULL values in an important column in the raw, unprocessed data is abnormally high. 

\begin{example}
When should I retrain my model?
\end{example}

The user performs a component history query on the inference component to visualize several metrics that track model staleness, such as the KL divergence between train and inference states, average prediction values, and median prediction values. They notice that it takes about a month for prediction quality to degrade enough to violate business SLAs. They encode a trigger to retrain the model monthly in the inference component \texttt{afterRun} method. 

%\aditya{factor out to not use SLO/SLI}

\begin{example}
Why is the accuracy much lower than expected right after deployment?
\end{example}

The user performs a cross-component query: they check the logs from the feature generation component that generated data for training and "propagate" these tests to the component that generates the online features. They see the production training component fails the new tests, indicating that there is a discrepancy between the online and offline feature generation code.

\begin{example}
Why are these clients complaining about the predictions we gave them over the last several months?
\end{example}

The user performs a cross-component history query: they "slice" the outputs, aggregate traces for outputs in the slice, and sort the \texttt{ComponentRun}s in the traces by frequency in descending order. They notice that the top-ranked \texttt{ComponentRun} is a preprocessing component that hasn't been refit in 6 weeks.

\mltrace{} provides a predefined set of commands, such as \texttt{trace}, \texttt{inspect}, and \texttt{history}, for easy query access. For more specific queries, users can query the logs and metadata via SQL. Many challenges stem from executing these queries quickly and presenting the results in a digestible way.

\section{Demo and Challenges}
\begin{figure*}
    \centering
    \vspace{-10pt}
    \begin{subfigure}[c]{0.49\linewidth}
    \fbox{\includegraphics[width=\linewidth]{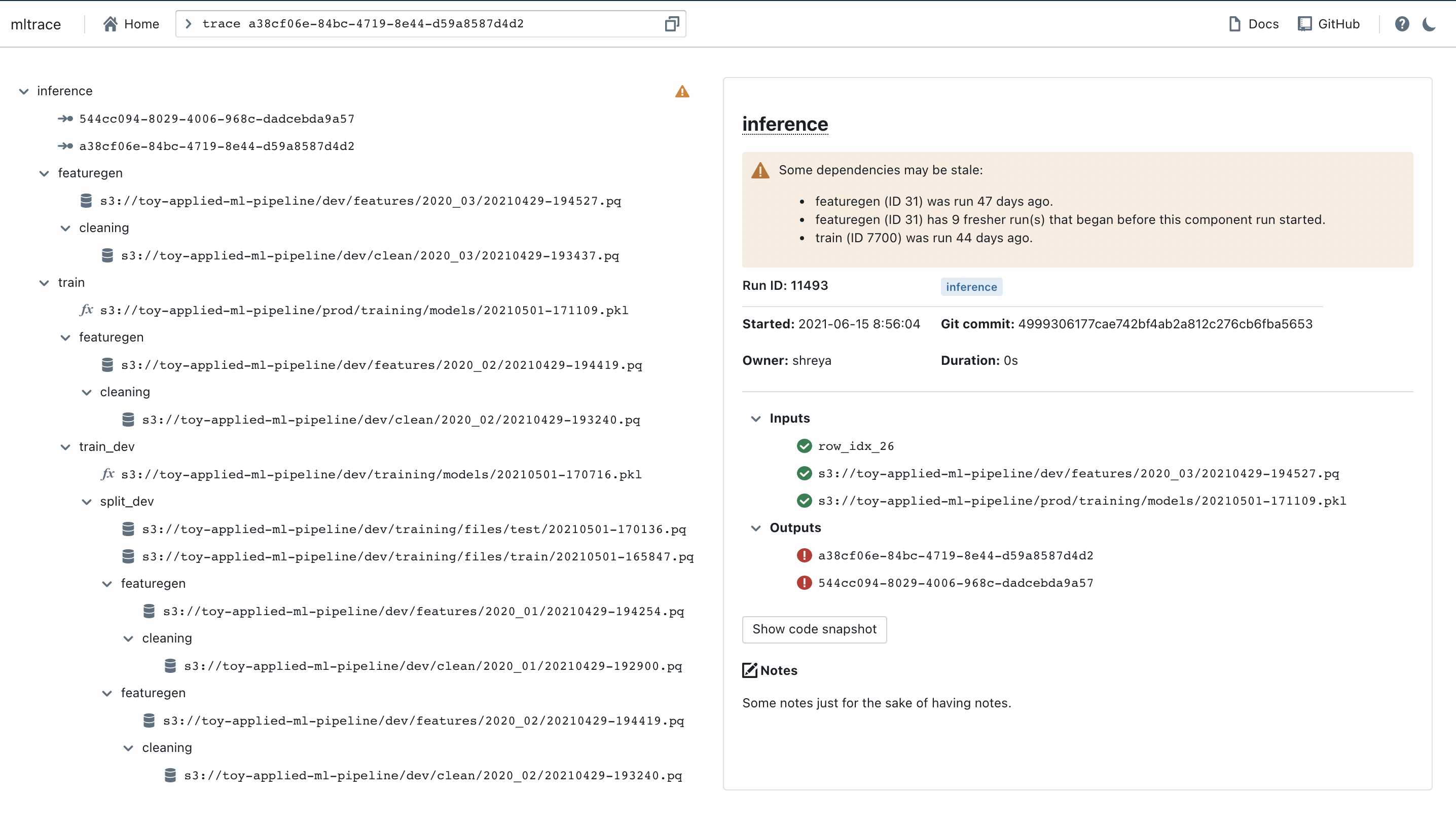}}
    \caption{\texttt{trace}}
    \end{subfigure}
    \hfill
    \begin{subfigure}[c]{0.49\linewidth}
    \fbox{\includegraphics[width=\linewidth]{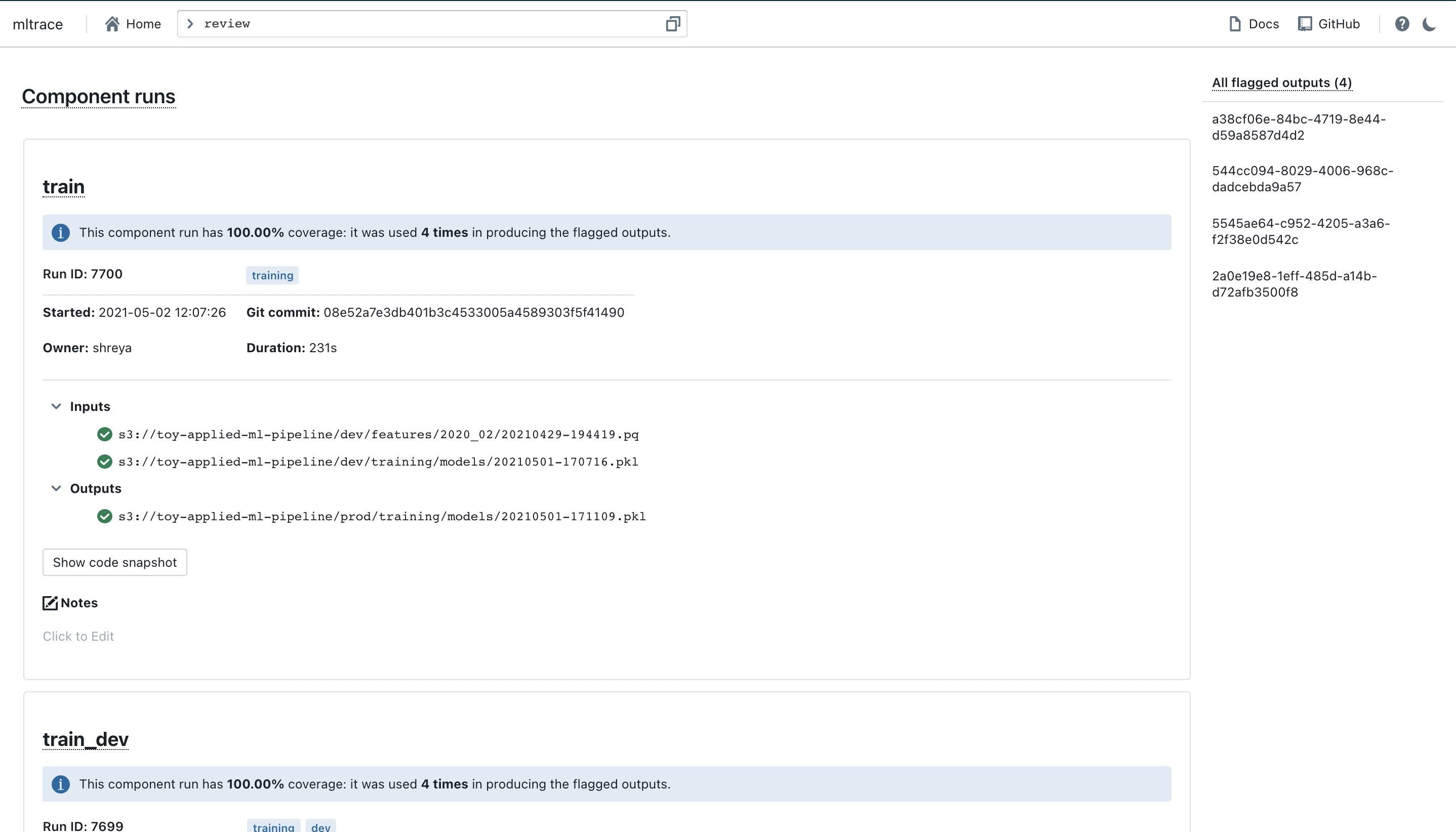}}
    \caption{\texttt{review}}
    \end{subfigure}
    \vspace{-5pt}
    \caption{Screenshots of the \mltrace{} web UI. In the \texttt{trace} command view, the red error icons correspond to problematic outputs that a user has manually flagged for review.}
    \label{fig:ui}
    % \vspace{-15pt}
\end{figure*}

% The current \mltrace{} release performs logging and tracks data flow to enable tracing. The version in development includes the \texttt{beforeRun} and \texttt{afterRun} trigger templates. The prototype only works with Python, but we are also working on gRPC integrations to be language-agnostic.
% \aditya{We've already described most
% of what's in the previous paragraph.
% Maybe just condense down to one line -- "We have already
% described our \mltrace prototype in Section X." -- and smush with next para.}

We have deployed \mltrace{} on an ML pipeline that predicts, for the NYC Taxicab dataset\footnote{\scriptsize \texttt{https://www1.nyc.gov/site/tlc/about/tlc-trip-record-data.page}}, whether a rider will give a high tip (at least 20\% of the fare) to a driver. there are currently 8 commands supported by the UI to determine history for a component, trace output IDs, and review faulty predictions to determine which components to debug. The \mltrace{} UI for this pipeline is  online\footnote{\scriptsize \texttt{https://bit.ly/3j1tqO8}}, and screenshots of the UI are shown in Figure~\ref{fig:ui}. 

Our \mltrace prototype makes small steps towards our
overall vision for ML observability. 
Next, we describe some relevant research challenges centered around logging, monitoring, and querying.

% \subsection{Logging Challenges}

\topic{Lightweight logging of intermediates} The current \mltrace{} prototype automatically logs inputs and outputs at component runtime. It quickly becomes expensive to log and store all intermediates of pipelines. One insight to reduce overhead is that users will only selectively query logged intermediates (e.g., inspect the head of a dataframe or summary statistics for a table). For each component, we can learn from query patterns over time to inform what goes into component run logs, thereby reducing logging latency and storage footprints.

\topic{Computing data distribution shift} Determining whether empirically observed groups of data points are significantly different is an unsolved algorithmic problem in the case where the shape of the distribution is unknown~\cite{sugiyama}. Computing simple metrics like the mean and median is a good start but can fail when skew and kurtosis changes. Computing well-known metrics like the Kolmogorov-Smirnov test statistic can be time-consuming and expensive (i.e., data needs to fit in memory) and produce too many false positive alerts~\cite{dataval}. We propose leveraging approximate query processing (AQP) techniques to operationalize computing distribution shift on large datasets. For instance, we can bin data points and perform resorvoir sampling to reduce latency while maintaining "good-enough" estimates for whether data has changed.

\topic{Determining when to retrain models} Efficiently computing distances between large-scale data distributions is only one step towards the goal of determining when to retrain models or refresh components (e.g., a preprocessor or cleaning criteria). We need to be able to track distribution shift along \emph{streams} of data, not just static datasets as previously explored in ML research communities~\cite{quinonero2009dataset, miller2020effect, subbaswamy2019preventing}. A challenge is to efficiently track distance metrics for various sliding window sizes and distill this information to accurately determine when components need to be refreshed.  Another research problem is to optimize  tracking shifts for groups of features or columns, which can combinatorially explode computation and storage.

\topic{Building easy-to-use interfaces for pipeline construction and debugging} Our library of predefined components needs to have both high coverage and accuracy for the kinds of tests and metrics users will want to compute and monitor. This requires research to understand reusable required tests and metrics across different ML-related tasks, such as robotics or recommender systems. Furthermore, our UI must help users holistically understand pipeline behavior. As described in Section~\ref{sec:sysexec}, pipeline DAGS could be large and complex, motivating new methods to draw human attention to summaries and anomalies (i.e., the most problematic components). Another challenge is to automatically determine where bugs lie in the pipeline so users do not have to sift through logs at every component run to address their own hypotheses.

\begin{full}
Our prototype makes small steps towards our
overall vision for ML observability. 
Next, we describe some remaining research challenges,
organized around the three dimensions of the system described in Section~\ref{sec:system}.

\topic{Developing a component library} The \texttt{beforeRun} and \texttt{afterRun} functions run frequently and may access and aggregate large amounts of state, making it hard to execute these functions quickly
(even if approximately).
We also need to ensure that the results of metric computations are \emph{accurate}: determining whether empirically observed groups of data points are significantly different is an unsolved algorithmic problem in the case where the shape of the distribution is unknown \cite{sugiyama}. Computing simple metrics like the mean and median is a good start but can fail when skew and kurtosis changes. Computing well-known distance metrics like Jensen-Shannon divergence or the Kolmogorov-Smirnov test can be expensive and produce too many false positive alerts \cite{dataval}. Finally, not only do we need to optimize for speed and correctness, but we also need to populate the library with useful components and functions for the practitioner to pick and choose from. This library should be complete, or cover most of the ways ML pipelines can fail.

\topic{Automatically inferred, lightweight logging} A key benefit of \mltrace{} is that it is lightweight and allows users to store data themselves. \aditya{what data are they storing?} This means that whenever a user runs components, \mltrace{} needs to (1) resolve component run dependencies without the user specifying them manually and (2) store copies of data and artifacts and deduplicate them on successive runs. Data versioning is well-studied, but extra challenges arise in storing models efficiently, especially when models are fine-tuned online during inference.
\aditya{Something like: As time progresses, we may want
to compact these logs to allow for aggregate queries
to still be performed, even if individual tracing
is no longer relevant on very old data---determining how to balance storage
and efficiency/utility is a challenge.}

\topic{Usable and expressive query interfaces} We want users to have fine-grained control in writing tests, building their pipelines, choosing what to monitor, and asking specific questions. Software observability queries are straightforward to execute, since the metrics are few and simple (i.e., mean response time for a REST endpoint). ML queries bring an unseen complexity to observability: for example, while debugging a pipeline, the user may want to aggregate over thousands of metrics over a window of a few months, which could be millions of data points. Furthermore, synthesizing large amounts of information to present to the user might also be challenging.\aditya{Say something about how the DAGs can get really complex---we need methods to draw human attention to summaries and anomalies (i.e., the most problematic components)?} It is hard for any human to visualize the variance of each feature for a model with thousands of features.

\topic{Regulation} Software observability systems typically focus on tracing a request "backwards" through all the steps involved in producing the response. Regulation such as GDPR requires ML application builders to be able to trace "forwards" -- for example, to identify and delete all artifacts derived from a user's data when a user deletes their account. Deleting these artifacts, which could include trained models, could break the ML application. Our solution would need to trigger relevant alerts to retrain models, recompute column bounds, and more to minimize ML application downtime. Naively recomputing artifacts any time a user deletes data could be computationally expensive.
\aditya{This is nice but can be possibly shortened to 2 lines and combined with either the logging or the interface challenges bit}
\end{full}

\begin{full}
\topic{Storing and materializing large, frequently-changing component states} A key benefit of \mltrace{} is that it is lightweight and allows users to store data themselves. This means that whenever a user runs components, \mltrace{} needs to store copies of data and artifacts and deduplicate them on successive runs. Data versioning is well-studied, but extra challenges arise in storing models efficiently, especially when models are fine-tuned online during inference. For the user to monitor inputs and outputs, \mltrace{} needs to store and materialize large states for each component at runtime, which would include all the I/O to each component over a user-defined period of time.

\topic{Efficiently and effectively computing metrics that measure component performance} Determining whether empirically observed groups of data points are significantly different is an unsolved algorithmic problem in the case where the shape of the distribution is unknown \cite{sugiyama}. Computing simple metrics like the mean and median is a good start but can fail when skew and kurtosis changes. Computing well-known distance metrics like Jensen-Shannon divergence or the Kolmogorov-Smirnov test can be expensive and produce too many false positive alerts \cite{dataval}. Second, even when the user knows what distance metrics they want to monitor, computing these metrics can be computationally expensive and redundant when two features represent similar concepts. We can leverage approximate query processing and batch queries to compute these distance metrics, but given the large number of metrics we need to regularly recompute, it will be necessary to eliminate redundancy in metric computation in order to alert the user when pipeline performance may be decreasing.

\topic{Fast query execution for debugging} Software observability queries are straightforward to execute, since the metrics are few and simple (i.e. mean). ML queries bring an unseen complexity to observability, as the user may want to aggregate over thousands of metrics over a window of a few months, which could be millions of data points. 

\topic{Fault tolerance} Many times, a user will want to rerun components because they found a bug, and these reruns could correspond to component runs that occurred months ago. Failed tests in the reruns need to trigger updates to the component runs that depended on the reruns. Metrics need to be recomputed not only for the reruns, but for component runs that lie within a window after the original execution times. Doing all of this efficiently might be challenging.

\topic{Interface} We want users to have fine-grained control in writing tests, building their pipelines, choosing what to monitor, and asking specific questions. We also want to provide a library of off-the-shelf components that users can easily string together and commands that answer common observability questions. Maintaining flexibility at this interface level might be challenging. Furthermore, synthesizing large amounts of information to present to the user might also be challenging.
\end{full}
\end{full}

\begin{full}
\section{Prototype}
% Current status (what we have implemented):

% \begin{itemize}
%     \item Logging I/O pathnames for components
%     \item Tracing outputs
%     \item \texttt{review} feature for people to flag problematic outputs and get suggestions for what components to look into first
%     \item Component staleness warnings (not data distribution specific)
%     \item UI
% \end{itemize}

% Next steps:

% \begin{itemize}
%     \item Make Python agnostic
%     \item Monitoring metrics computed on output values to detect distribution shift
%     \item "Triggers" at the I/O level
%     \item Lower latency when logging (show how slow the decorator is)
%     \item Allow user to log I/O itself, not just the path (but how to do this efficiently?)
%     \item Allow users to run their own queries on the logs, not just our defined commands
%     \item Finer grained logging within the component level
% \end{itemize}

\mltrace{} is separated into client-side and server-side parts. The server-side part stores the logs. On the client side, a user can set their environment variable to the server address or call a function to connect to the server before creating and sending any logs via \mltrace{} logging functions. To query, the user interacts with \mltrace{} via a set of commands we have defined. We will grow this set as we add more functionality. We currently offer a web UI and a CLI for users to run commands.

\subsection{Current}

The \mltrace{} prototype supports only Python-based workflows. Table \ref{tab:commands} lists the commands that \mltrace{} currently supports. Several commands represent straightforward \texttt{SELECT *}s from the logs. The \texttt{trace} command displays the end to end data flow for an output. To trace the data flow for an output, \mltrace{} simply performs a depth-first search on the ComponentRun graph to compute all the ComponentRuns involved in producing the output. The \mltrace{} prototype supports the two notions of staleness defined in section \ref{sec:dataflow}.  

\begin{table}[h]
    \centering
    \begin{tabularx}{\columnwidth}{|X|X|X|}
    \hline
        \bf Command & \bf Description \\ \hline
        \texttt{recent} & Shows recent component runs, also the home page \\ \hline
        \texttt{history COMPONENT\_NAME} & Shows history of runs for the component name. Defaults to 10 runs. Can specify number of runs by appending a positive integer to the command, like `history etl 15` \\ \hline
        \texttt{inspect COMPONENT\_RUN\_ID} & Shows info for that component run ID \\ \hline
        \texttt{trace OUTPUT\_ID} & Shows a trace of steps for the output ID \\ \hline
        \texttt{tag TAG\_NAME} & Shows all components with the tag name \\ \hline
        \texttt{flag OUTPUT\_ID} & Flags an output ID for further review. Necessary to see any results from the `review` command. \\ \hline
        \texttt{unflag OUTPUT\_ID} & Unflags an output ID. Removes this output ID from any results from the `review` command. \\ \hline
        \texttt{review} & Shows a list of output IDs flagged for review and the common component runs involved in producing the output IDs. The component runs are sorted from most frequently occurring to least frequently occurring. \\ \hline
    \end{tabularx}
    \caption{Current \mltrace{} command list}
    \label{tab:commands}
\end{table}

Note that the user currently needs to define outputs for every ComponentRun. This is fine when several components are intermediate stages in producing the final result of the ML pipeline. When the component is an endpoint, however, the user needs to create output IDs corresponding to the outputs of the endpoint. \mltrace{} has a utility function to create any number of random IDs.

The \texttt{flag}, \texttt{unflag}, and \texttt{review} commands are \mltrace{}'s initial attempt to give the user suggestions for where to begin debugging given faulty outputs. After an output is produced, users have the option to flag or unflag it at any time using the relevant command. The \texttt{review} command computes and collates the traces for each flagged output, orders the ComponentRuns in the traces from most frequently to least frequently used, and returns this list of ComponentRuns to the user. The idea here is that the most frequently used ComponentRuns may have errors and could be a good place to start debugging. \shreya{do i need to write an algorithm here}

\subsection{Integration example}

\subsection{Roadmap}

\mltrace{} has a long ways to go to meet all the requirements defined in table \ref{tab:userreqs} and a difficult journey ahead, as discussed in section \ref{sec:challenges}. To achieve full observability, we need to support the proactive user requirements. Most immediately, we need to store the data. We then plan to develop mechanisms to efficiently measure and monitor values over time. This will involve extending our ComponentRun abstraction to support constraints and triggers, where we can encode checks for data distribution shifts (PRO1) and basic "unit" tests for data science work (PRO3).

We also need to extend our logging mechanisms beyond Python. One option is to use gRPC for logging since in supports multiple languages. We also need to support asynchronous logging, since the current decorator approach can have significant overhead when the number of output IDs for a component is greater than 1000. Finally, we need to explore finer-grained logging and better integrations with component-specific tools. We are actively thinking about solutions to all of these issues.

\shreya{should we include an example of using mltrace}
\end{full}

\section{Conclusion}
We proposed new research challenges in ML observability through a taxonomy of detecting, diagnosing, and reacting to ML bugs, \mr{helping make sense of performance with incomplete information, 
identify potential issues in ML pipeline components, and trace the source of errors}. We discussed a high-level architecture of a bolt-on ML observability system\mr{, and introduced our prototype, \mltrace{}}\techreport{, a lightweight, platform-agnostic end-to-end observability tool for ML applications}. 
We call on the database community to contribute
to the vision of ML observability\techreport{, 
supporting users who are comfortable with their existing toolstack, while}\papertext{ by} alleviating the many
the data management and querying concerns that come with production ML.

% \mltrace{} is still under active development: we are working on testing, monitoring, and alerting mechanisms as well as populating a component library for practitioners to construct ML pipelines off-the-shelf.

%\section{References}
%\printbibliography
\bibliographystyle{IEEEtran}
% \scriptsize
\bibliography{main}
\end{document}